\documentclass[aps,superscriptaddress,twocolumn]{revtex4-2}
\usepackage[utf8]{inputenc}
\usepackage{amsmath,amssymb}
\usepackage{graphicx}
\usepackage{booktabs}
\usepackage{xcolor, soul}
\usepackage{hyperref}
\usepackage[capitalise]{cleveref}%To do notes
\usepackage{todonotes}
\usepackage{braket}
\usepackage{threeparttable}

\begin{document}
%\title{A high-performance cryogenic piezoelectric resonator based on \\ thin-film barium titanate} 
%\title{Thin-film barium titanate for cryogenic quantum acoustics}
%\title{Cryogenic piezoelectricity in thin-film barium titanate}
\title{Piezoelectric resonators in thin-film barium titanate from room temperature to millikelvin}

%Placeholder title
\date{\today}
\author{Hao Tian}
\thanks{These authors contributed equally.}
\email{haotian@caltech.edu}
\affiliation{The Gordon and Betty Moore Laboratory of Engineering, California Institute of Technology, Pasadena, California 91125}
\affiliation{Institute for Quantum Information and Matter, California Institute of Technology, Pasadena, California 91125}
\author{Shu-Yuan Chang}
\thanks{These authors contributed equally.}
\affiliation{The Gordon and Betty Moore Laboratory of Engineering, California Institute of Technology, Pasadena, California 91125}
\affiliation{Institute for Quantum Information and Matter, California Institute of Technology, Pasadena, California 91125}
\author{Nuha Akhtar}
\affiliation{Department of Mechanical Engineering, University of California, Berkeley, California 94720 }
\author{Kasra Sardashti}
\affiliation{Department of Physics, University of Maryland, College Park, MD 20742}
\affiliation{LPS Qubit Collaboratory, College Park, MD 20740}
\author{Mohammad Mirhosseini}
%\homepage{http://qubit.caltech.edu}
\affiliation{The Gordon and Betty Moore Laboratory of Engineering, California Institute of Technology, Pasadena, California 91125}
\affiliation{Institute for Quantum Information and Matter, California Institute of Technology, Pasadena, California 91125}

\begin{abstract}
Ferroelectric materials, with their strong nonlinearities, underpin key technologies across radio-frequency (RF) signal processing, optical communications, and emerging quantum systems. Barium titanate (BTO) is a notable example, combining strong piezoelectric and electro-optic responses. While bulk BTO has been studied for decades, the piezoelectric properties of its recently available thin films, and their behavior at the millikelvin temperatures relevant to quantum hardware, remain largely unexplored. Here, we fabricate and characterize surface acoustic wave (SAW) resonators on thin-film BTO. The measured devices exhibit high electromechanical coupling $(k^2_{\text{eff}}\approx 14\%$ at 5.2 GHz) and operate up to 7.8 GHz. From these measurements, combined with finite-element modeling of the multi-domain microstructure, we extract an effective piezoelectric coefficient $d_{33,\text{eff}}$ of 53 pC/N, comparable to bulk BTO. Exploiting the intrinsic ferroelectricity, we further demonstrate low-voltage switching with a fast ($\sim$100 ns) response, attractive for reconfigurable RF front-ends and parametric amplifiers. Extending these measurements to millikelvin temperatures, we find that the piezoelectric response persists, with $d_{33,\text{eff}}\approx$ 19 pC/N, pointing to the potential of BTO for piezoelectric coupling in superconducting quantum circuits. These results position thin-film BTO as a promising piezoelectric platform for both classical and quantum information technologies.
%Numerical modeling of the multi-domain microstructure yields an effective piezoelectric coefficient $d_{33,\text{eff}}$ of 53 pC/N, comparable to bulk BTO, enabling resonators with high electromechanical coupling $(k^2_{\text{eff}}\approx 14\%$ at 5.2 GHz) and operation up to 7.8 GHz. The intrinsic ferroelectricity further enables low-voltage switching with a fast (
%$\sim$100 ns) response, attractive for reconfigurable RF front-ends. Extending these measurements to millikelvin temperatures, we find that the piezoelectric response persists, with $d_{33,\text{eff}}\approx$19 pC/N, pointing to the potential of BTO for piezoelectric coupling in superconducting quantum circuits. These results position thin-film BTO as a promising piezoelectric platform for both classical and quantum information technologies.
\end{abstract}

\maketitle

\section*{Introduction}
Ferroelectric materials have been widely used in modern electronic and RF technologies, owing to their high dielectric constants and large nonlinearities that arise from strong spontaneous polarization \cite{setter2006ferroelectric}. More recently, advances in synthesis, characterization, and nanofabrication of ferroelectric thin-films have facilitated large-scale integration of devices.  Despite this progress, the ever-growing demand for lower energy consumption, wider operation bandwidth, and smaller device footprint continues to motivate the search for materials with stronger nonlinear response.
%high frequency, high bandwidth, switchable --> high performance --> need large k^2 and piezo
%Acoustic resonators made from ferroelectric thin films have become the pivot in today's wireless communication systems by working as RF filters with compact size and low insertion loss.
In the case of piezoelectric devices, a stronger electromechanical coupling, higher frequency, and intrinsically switchable response are actively pursued for acoustic filters in wireless communication networks \cite{du2024near, koohi2020reconfigurable, anderson2026high}. In parallel, the growing research on quantum computing has opened a new domain of applications for these materials in quantum transduction. For instance, piezo-optomechanical transducers are sought to build optical interconnects for superconducting qubits \cite{han2021microwave}. Piezoelectric coupling of mechanical resonators to superconducting qubits is also explored as a mean of realizing quantum hardware components for encoding bosonic qubits \cite{von2024engineering} or quantum random access memories for quantum computing \cite{hann2019hardware}. Physical implementation of these systems requires a strong electromechanical response at millikelvin temperatures and minimal acoustic loss. Previous works have demonstrated proof-of-concept devices using a range of piezoelectric materials, including AlN \cite{von2024engineering}, GaAs \cite{forsch2020microwave}, and $\mathrm{LiNbO}_3$ \cite{wollack2022quantum}, though practical applications will require further improvements in material-limited coupling strength and loss.
%While these experiments have achieved strong electromechanical coupling to qubits, material-limited coupling strength and loss constrain current systems proof-of-concept operation rather than practical applications, motivating the exploration of platforms with stronger nonlinearities. 

Barium titanate (BaTiO$_3$, BTO), a classic perovskite ferroelectric discovered in the 1940s \cite{von1946high}, has long been studied for its strong piezoelectric and electro-optic properties, owing to its relatively low Curie temperature and the resulting strong spontaneous polarization near the phase transition \cite{anderson2025quantum, zgonik1994dielectric}. As a lead-free material, BTO also offers an environmentally friendly alternative to traditional ferroelectrics such as lead zirconate titanate (PZT).  Interest in BTO for integrated photonics has grown only recently, driven by advances in growing single-crystal BTO thin films. Electro-optic modulators have been demonstrated with large electro-optic coefficients and data rates exceeding 250 Gb/s \cite{abel2019large, eltes2023thin}, as well as cryogenic operation that retains large electro-optic responses \cite{eltes2020integrated}. 
Despite this rapid progress in BTO photonics, its piezoelectric properties have received comparatively limited attention. Although MHz- and GHz-range BTO resonators have been demonstrated \cite{lee2022ultrathin, lee2013intrinsically, anderson2026high}, the mechanical and piezoelectric properties of BTO thin films, particularly at cryogenic temperatures, remain comparatively underexplored \cite{anderson2025quantum}.

In this work, we use surface acoustic waves (SAW) generated by integrated interdigital transducers (IDT) to probe the properties of BTO thin films. From SAW resonances and $k_{\text{eff}}^2$, we extract the mechanical compliance and piezoelectric coefficients, finding values comparable to bulk BTO. An effective piezoelectric coefficient $d_{33,\text{eff}}$ of 53$\pm$5 pC/N is obtained, enabling acoustic resonators whose electromechanical coupling and operating frequency compare favorably with state-of-the-art devices. Furthermore, the intrinsic ferroelectricity of BTO provides in-situ switching and frequency tuning with a fast switching time of $\sim$100 ns and a low switching voltage compatible with mobile electronic devices.
Cryogenic operation of BTO SAWs at millikelvin temperatures shows that the piezoelectric response persists, pointing to the potential of BTO for piezoelectric coupling in superconducting quantum circuits. Together, these results introduce thin-film BTO as a promising piezoelectric platform for classical and quantum information technologies, ranging from reconfigurable RF transceivers to quantum transducers and memories.
%These results support advanced applications of BTO thin films in next-generation 5G/6G technologies, reconfigurable RF systems, and quantum information processing.

%high efficiency, low-cost, high frequency or speed, large bandwidth
%compare to LN, BTO is tunable, larger piezo at both RT and cryo. 
\begin{figure*}[ht]
    \centering
    \includegraphics[width=\textwidth]{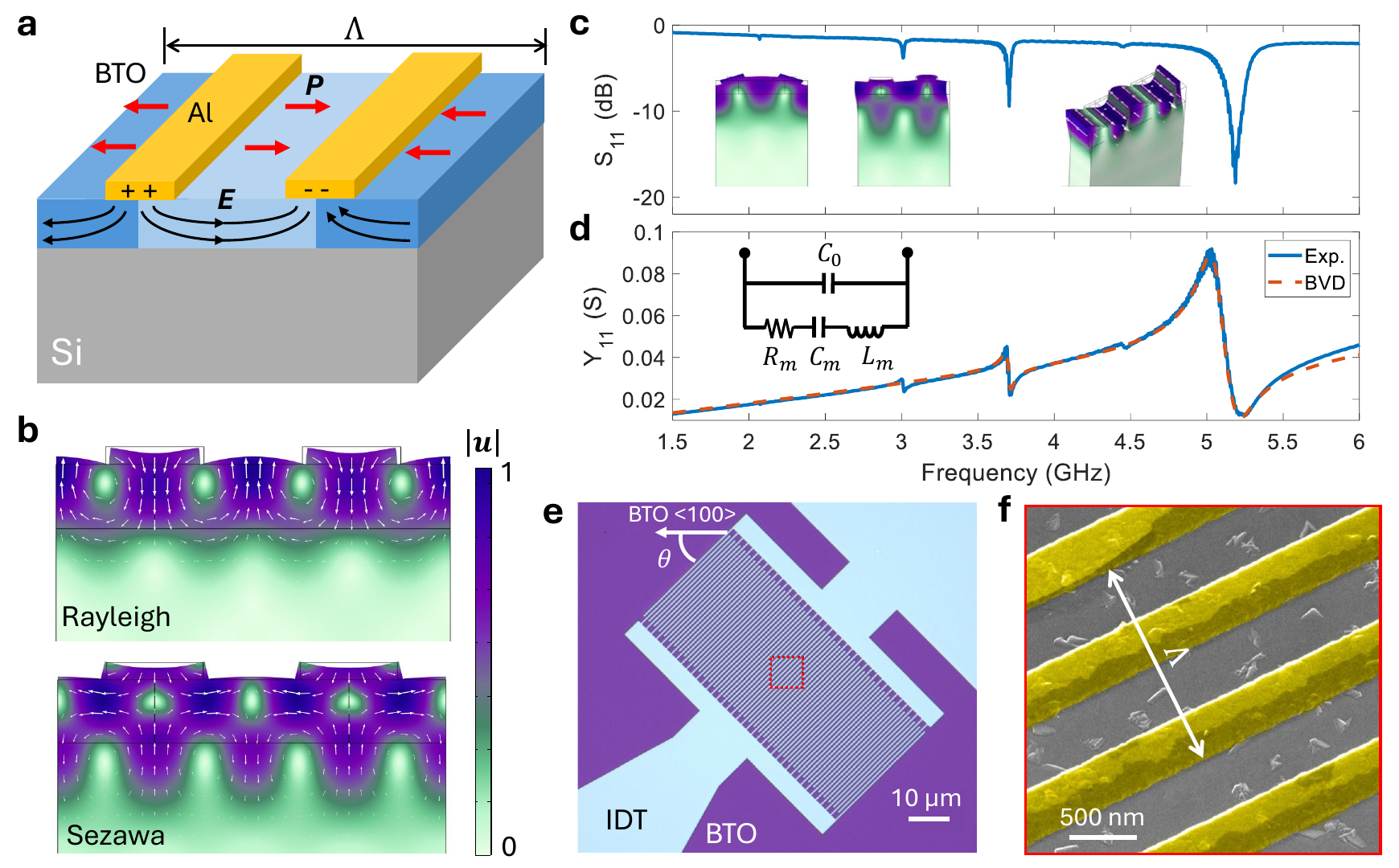}
    \caption{\textbf{Device design and characterization.} \textbf{a,} Three-dimensional (3D) schematic showing one period $\Lambda$ of the IDT, with aluminum (yellow) electrodes. The black and red arrows denote the electric field and polarization in BTO, respectively. \textbf{b,} Numerical simulation of higher-order Rayleigh (top) and Sezawa (bottom) modes matched with the periodically poled IDT. The colormap is the normalized displacement magnitude $\textbf{u}$ with white arrows denoting the displacement direction. \textbf{c,} Measured microwave reflection spectrum ($S_{11}$) of a typical IDT under 30 V bias. Insets from left to right are mode profiles for the fundamental Rayleigh and Sezawa modes, and the shear horizontal mode. \textbf{d,} Calibrated admittance ($Y_{11}$) of the IDT (blue solid line) with BVD model fitting (dashed red line). The inset shows the equivalent circuit of the BVD model. \textbf{e,} Optical microscope of the fabricated IDT. The IDT orientation $\theta$ is defined with respect to the BTO $\langle100\rangle$ direction (or $\langle001\rangle$ due to multi-domain structure, see later section).  \textbf{f,} False-colored SEM of the zoom-in region in red dashed box in \textbf{e}. }
    \label{fig1} 
\end{figure*}

\section*{Device design}
The development of epitaxial single-crystal SrTiO$_3$ (STO) on silicon has facilitated the growth of high-quality BTO films by providing a lattice-matched buffer layer \cite{posadas2021thick}. In this work, we use a commercially available stack of 300 nm RF-sputtered BTO on a 280 $\mu$m STO-buffered Si substrate (from La Luce Cristallina). 
At room temperature, BTO is stabilized at the tetragonal ferroelectric phase (4$mm$) with spontaneous polarization parallel to the c-axis (long axis) \cite{acosta2017batio3, li2006temperature}. 
However, previous studies have shown that thin-film BTO contains random domains with opposite polarizations \cite{eltes2020integrated, abel2019large}.  
As a result, the macroscopic polarization averages to zero, suppressing the piezoelectric response. Poling with an external electric field is therefore needed to align the domain polarizations \cite{geler2022ferroelectric, nordlander2020}.
To allow convenient poling with in-plane electrodes, we use an a-axis oriented BTO film (referred to as a-domain), in which the c-axis (thus the polarization) lies within the film's plane.    

We use SAW to study the BTO thin film because the surface-confined acoustic field overlaps well with the film. It is conventionally generated by IDTs, where an alternating electric field from interleaved electrodes drives piezoelectric contraction and expansion, launching an acoustic wave along the surface \cite{du2024near}.
Figure~\ref{fig1}a shows one period ($\Lambda$) of the IDT. The poling is achieved by applying a direct current (DC) voltage to the electrode pair, where the alternating electric field (black arrows) aligns the polarization (red arrows) and forms a periodic poling structure. SAW is excited through an alternating current (AC) voltage signal at microwave frequencies. Unlike conventional IDTs, this periodically poled design excites higher harmonic acoustic waves with wavelength of $\lambda=\Lambda/2$ because polarizations between electrodes share the same polarity relative to the electric field, resulting in a piezoelectric response with $\Lambda/2$ period.
An advantage of this design is that it reaches higher frequencies for the same IDT dimensions, easing the nanofabrication constraints that limit frequency scaling in conventional SAW resonators \cite{du2024near}. 

Figure \ref{fig1}b shows the numerical simulation of the second harmonic of two families of SAW modes, namely the Rayleigh and Sezawa modes. 
They differ in that the displacement in BTO is primarily transverse for Rayleigh mode and longitudinal for the Sezawa mode. 
The number of finger pairs and the electrode aperture length affect mode confinement, electromechanical coupling, and impedance matching, which are studied in Appendix C.  Unless otherwise noted, we present results for IDTs with 1.8 $\mu$m pitch, 40 finger pairs, and 30 $\mu$m aperture. The electrodes are made from 60 nm aluminum (Al) patterned by a standard lift-off process. An optical microscope image of a typical IDT is shown in Fig. \ref{fig1}e. A scanning electron microscope (SEM) image in Fig. \ref{fig1}f shows the electrodes (yellow) and the BTO surface (gray). We attribute the pyramid-like protrusions on the BTO surface to c-axis domains and random polycrystalline regions (see Appendix B) \cite{posadas2021thick}. 

\section*{Characterization of BTO SAW}
The BTO SAW is characterized by measuring the microwave reflection spectrum ($S_{11}(\omega)$) from the IDT using a vector network analyzer under a persistent 30 V bias that fully poles the film; the corresponding poling hysteresis and switching behavior are presented in the later section. A representative $S_{11}$ spectrum is shown in Fig. \ref{fig1}c, featuring two dominant resonances at 3.7 and 5.2 GHz, corresponding to the higher-order Rayleigh and Sezawa modes, respectively. We attribute the two weaker resonances at 2.1 and 3 GHz to the fundamental Rayleigh and Sezawa modes, respectively, based on their frequencies. Although these modes are expected to be uncoupled from the periodically poled IDT due to symmetry in the ideal case, slight imbalances in domain populations (and the resulting net polarization) produce nonzero coupling in practice. For simplicity, `Rayleigh' and `Sezawa' refer to the higher order modes in the following sections.
A shallow dip at 4.5 GHz corresponds to the shear horizontal (SH) mode, associated with the piezoelectric coefficient $d_{15}$. 
The Sezawa mode shows the strongest response because the in-plane strain and electric field align with the large $d_{33}$ coefficient of BTO \cite{zgonik1994dielectric}. This strong coupling produces an overcoupled resonance with intrinsic ($\gamma_i$) and external ($\gamma_e$) coupling rates of 76 and 113 MHz, respectively. We attribute the relatively low intrinsic mechanical quality factor $Q_i$ ($\approx$70) to substantial substrate radiation, evidenced by the coupling to bulk acoustic wave (BAW) modes of the Si substrate (see Appendix A and later sections). 

 %The intrinsic capacity of a piezoelectric to convert energy between the electrical and the mechanical domain is defined by the material electromechanical coupling coefficient $K^2= \frac{d^2}{s^E\varepsilon^T}$, with compliance $s^E$ and permittivity $\varepsilon^T$ taken at constant electric field and constant stress respectively. The corresponding device-level figure of merit is the effective electromechanical coefficient $k_{\text{eff}}^2$, which reflects how efficiently a particular device exploits the intrinsic material coupling. 
 
To quantify the piezoelectric strength, we use the effective electromechanical coupling coefficient ($k_{\text{eff}}^2$), which is a measure of how efficiently a piezoelectric material converts energy between electrical and mechanical forms (see methods). We extract $k_{\text{eff}}^2$ from the measured admittance ($Y_{11}$) of the IDT (see methods), as shown in Fig. \ref{fig1}d. The electrical response is modeled by an equivalent circuit with lumped elements, the widely used Butterworth-Van Dyke (BVD) model (see Methods) \cite{blesin2021quantum}.
Measuring multiple devices of the same design across the chip yields a maximum $k_{\text{eff}}^2$ of 8\% and 1\% for Sezawa and Rayleigh modes, respectively (see Appendix B). A detailed study of the IDT length and pair number yields a maximum $k_{\text{eff}}^2$ of 14.2\% at 5.2 GHz. 
We also explore frequency scaling (see Appendix C), reaching a maximum of 7.8 GHz while maintaining an electromechanical coupling of 6.2\%. These results compare favorably with state-of-the-art SAW resonators based on widely studied piezoelectric materials such as AlN, AlScN, ZnO, and GaN \cite{hadj2019sezawa, du2024near, ahmed2023super} (see Appendix D). 
%Although the $k_{\text{eff}}^2$ is still less than those based on $\mathrm{LiNbO}_3$ platform \cite{zheng2023near}, the CMOS-compatibility of BTO allows simpler fabrication and lower cost. 

%why high frequency and k2 is important for wireless communication, 5G, IoT, sensor: higher data rate, bandwidth
%where do we want to mention the device variation 
%HT: we want to say that the high frequency and k2 is demanded in today RF technologies, and our BTO SAW out performs most of the state of the art SAW resonators such as AlN, AlScN, ZnO, GaN, PZT with higher k2 and frequency. we can put a comparison table or figure in the supplement. To show that, we will need to give detail study of k2 on IDT pair and length, and give the frequency scaling study to show the high frequency operation, probably all in supplement?
%the results are preliminary and design is sub-optimal which however shows promising behariour and pave the path for high performnace RF devices in the future with improved design and fabrication. 
%although the k2 is still worse than LN, the fabrication is CMOS-compatible, more importantly, it is switchable and tunable . 

\begin{figure*}[ht]
    \centering
    \includegraphics[width=\textwidth]{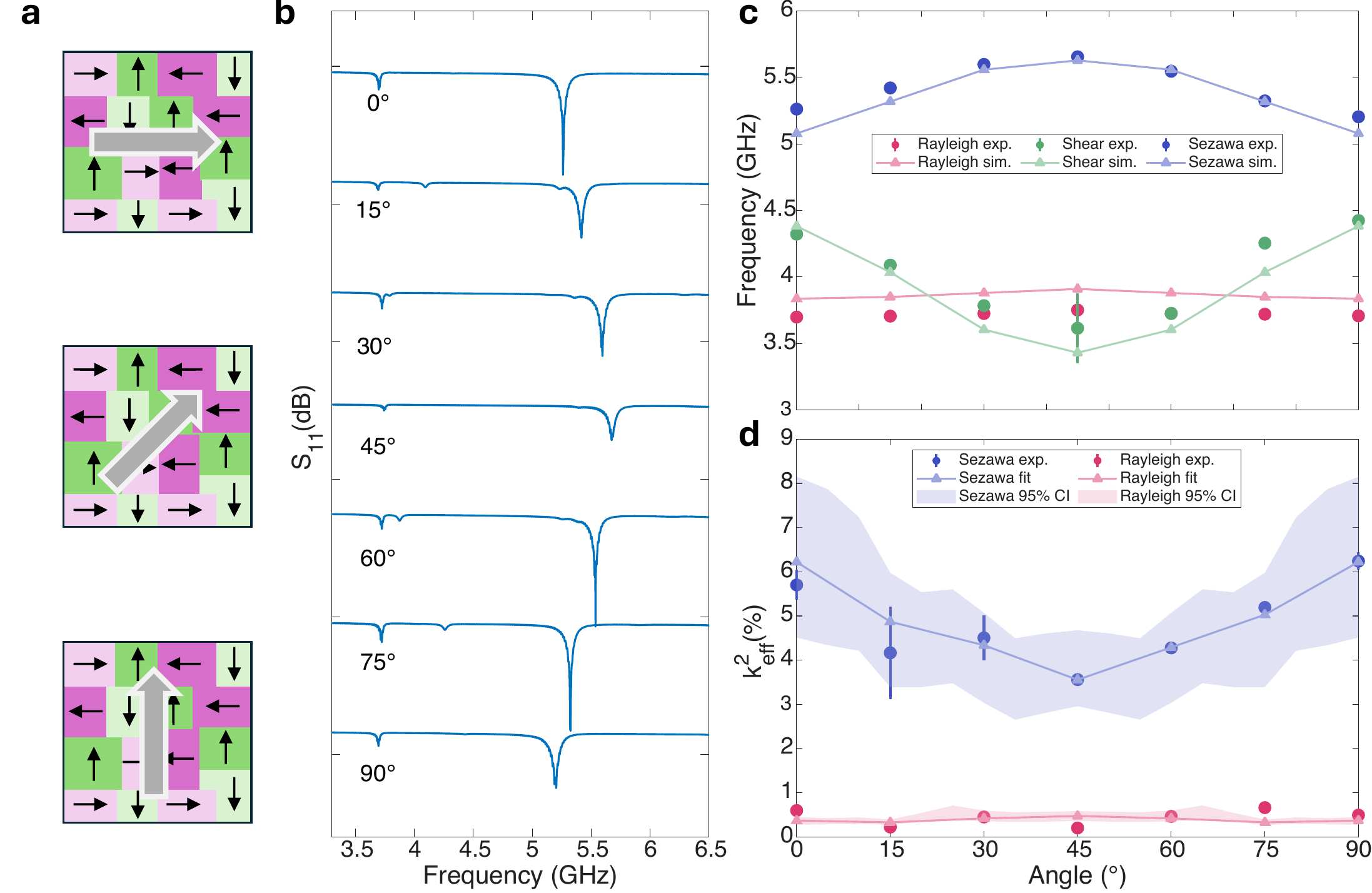}
    \caption{\textbf{Orientation dependence of the BTO SAW resonators.} \textbf{a,} Schematics of the multi-domain BTO model. The elastic and piezoelectric properties of the thin film is modeled as a mosaic ensemble of single domains with four orthogonal polarizations. The IDT is rotated relative to the BTO lattice from 0$^{\circ}$ to 90$^{\circ}$ (top to bottom). The black and gray arrows represent polarizations of each domain and the direction of IDT, respectively. \textbf{b,} $S_{11}$ spectra under different IDT orientations. Traces are offset vertically for better visibility, with spacing between y-ticks of 50 dB. Angle dependence of \textbf{c,} the mechanical frequencies and \textbf{d,} $k_{\text{eff}}^2$ for Rayleigh (red), shear (green), and Sezawa (blue) modes. Comparison between the measurement (circle) and FEM simulations (triangle) shows good agreement. The experimental error bars in \textbf{c} and \textbf{d} are the standard deviation from multiple measurements, with most of them smaller than the symbol. Shaded regions correspond to the 95\% CI of the extracted piezoelectric coefficients. }
    \label{fig:orientation} 
\end{figure*}

\section*{Extraction of piezoelectric coefficient}
In this section, the mechanical and piezoelectric properties of the BTO thin film and its crystallographic anisotropy are studied by rotating IDTs relative to the BTO film. The key material coefficients are extracted by analyzing the measured SAW responses with finite-element method (FEM) fitting. Previous studies have shown that, for epitaxial a-axis oriented BTO films, two orthogonal and energetically degenerate in-plane c-axes are supported, resulting in four polarization directions \cite{geler2022ferroelectric}, as shown in Fig. \ref{fig:orientation}a. Owing to the four-fold degeneracy of domain orientations, the IDT response exhibits a 90$^{\circ}$ periodicity (see Appendix H). The spectra of IDTs rotated from 0$^{\circ}$ to 90$^{\circ}$ are shown in Fig. \ref{fig:orientation}b. For BTO film in the tetragonal phase, the domain morphology can be modeled as a two-dimensional (2-D) mosaic pattern with randomly and equally distributed domain orientations \cite{li2006temperature}. Note that, despite the four possible orientations, a domain can only be switched by 180$^{\circ}$ \cite{nordlander2020}. As the domain sizes (50-100 nm \cite{reynaud2025enhancement}) are much smaller than the IDT period, the measured response reflects effectively homogenized elastic and piezoelectric constants averaged over the ensemble of these nano-domains. In the following, we develop models to calculate these effective macroscopic constants from those of individual domains.

We first extract the elastic parameters by matching the measured resonance frequencies to FEM acoustic simulations, as shown in Fig. \ref{fig:orientation}c. %The effective elastic tensor of the multi-domain BTO film is modeled by volume-averaging the rotated tensors from each domain, weighted by their populations \cite{hill1952elastic}. It provides a computationally efficient approximation that agrees well with more sophisticated models \cite{li2000effective}. 
%We computed the effective elastic properties of the multi-domain BTO film by volume-averaging the elastic/compliance tensors of each single domain population using the Voigt-Reuss-Hill (VRH) approximation \cite{hill1952elastic}. This volume-based approach is shown to be consistent with more complex self-consistent methods \cite{li2000effective} and with our FEM-based simulations (see Appendix I).  
We compute the effective elastic properties of the multi-domain BTO film using the Voigt-Reuss-Hill (VRH) approximation \cite{hill1952elastic}, averaging the stiffness and compliance tensors of the constituent domain variants over their relative populations. This computationally efficient homogenization approach agrees well with both self-consistent effective-medium models \cite{li2000effective} and FEM-based simulations (see Appendix I).  
Notably, using bulk BTO elastic constants from Ref. \cite{zgonik1994dielectric} as the input for each single domain, the simulated Rayleigh and Sezawa resonant frequencies agree reasonably with experiment, indicating that the mechanical properties are similar to those of bulk BTO. We do, however, observe a deviation in the measured shear mode frequency, which comes from the discrepancy in shear compliance $s_{44}$. 
Detailed modeling suggests this may arise from crystal-orientation disorder, such as mixed c-domains and polycrystalline regions (see Appendix I) \cite{posadas2021thick}.
%which is expected to affect $s_{44}$ more strongly than the other compliance components, since it is both the largest in magnitude and the most orientation-dependent.

\begin{figure*}[ht]
    \centering
    \includegraphics[width=15.5cm]{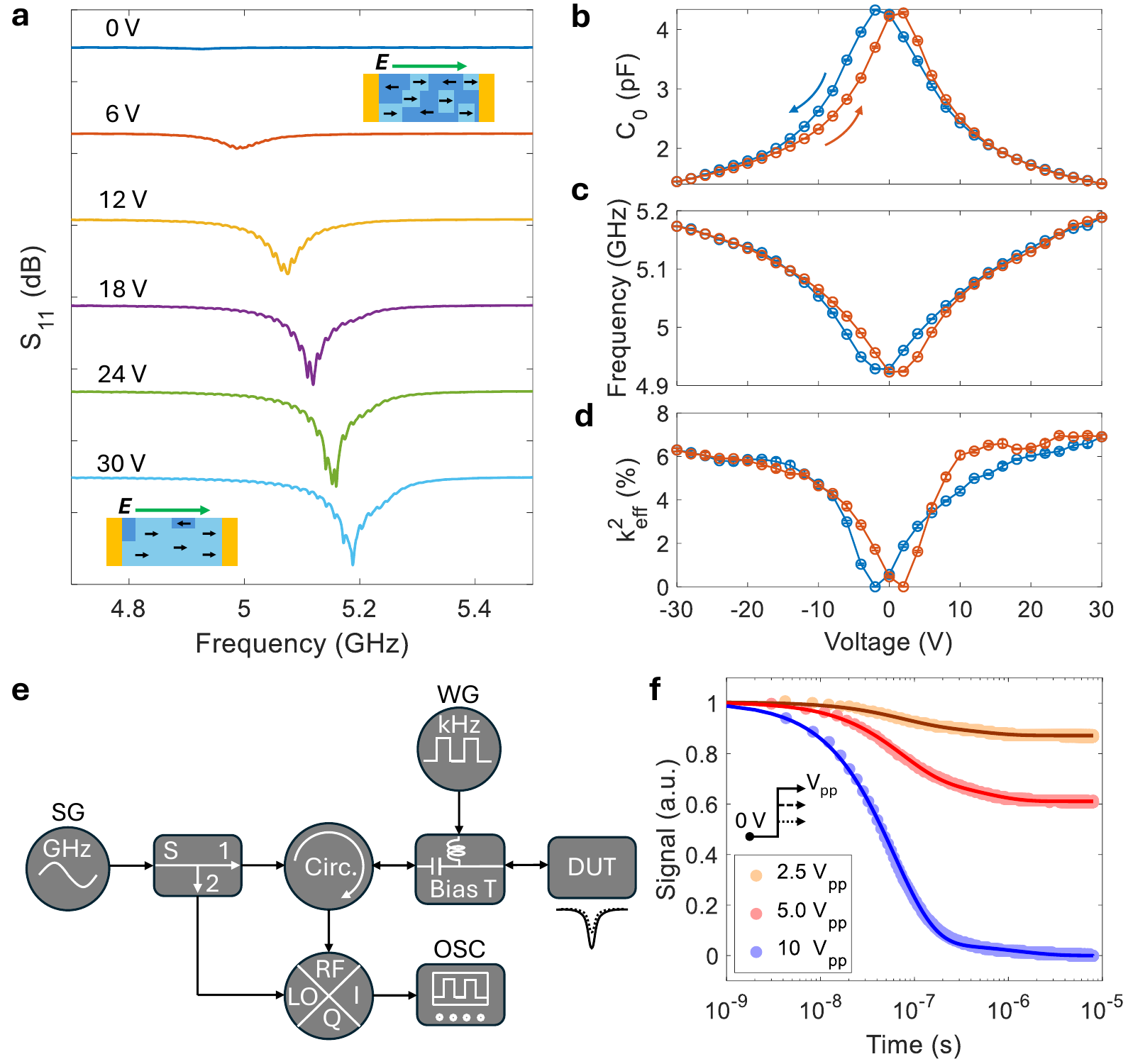}
    \caption{\textbf{Reconfigurable BTO SAW resonator.} \textbf{a,} $S_{11}$ response of the Sezawa mode with an increasing biasing voltage from 0 to 30 V (top to bottom). Each trace is vertically shifted for better illustration. The spacing between y-ticks is 20 dB. The insets schematically show the initial and final polarization states after aligning with the external electric field. \textbf{b-d,} Hysteresis curves measured with increasing (red) and decreasing (blue) DC bias for intrinsic capacitance $C_0$, resonant frequency, and electromechanical coupling $k_{\text{eff}}^2$, respectively. The error bar is the standard deviation from the measurements of four consecutive loops. \textbf{e,} Schematic of the setup for measuring the switching dynamics of the SAW resonator. A square wave (10 kHz) from the waveform generator (WG) switches the SAW resonance between two states. The magnitude of the reflected RF signal from the device is monitored through a homodyne detection using an IQ mixer. SG: signal generator, Circ: RF circulator, LO: local oscillator, OSC: oscilloscope, DUT: device under testing. \textbf{f,} Dynamics of the state switching when the bias voltage changes from 0 V to various peak-to-peak voltage $V_{\text{pp}}$, as illustrated in the inset. The homodyne signal is normalized such that the $V_{\text{pp}}=10$ V signal spans between 0 and 1. The solid lines are empirical two-time-constant exponential fittings.}
    \label{fig:switching} 
\end{figure*}

%The extension of the same method to the piezoelectric coefficients is nontrivial and generally not valid for complex multi-domain architecture.

For the piezoelectric coefficients, simple volume averaging is not directly applicable, because the elasto-electrical coupling must satisfy both the mechanical and electrical boundary conditions between adjacent domains.  We therefore construct an FEM model of the randomized mosaic domain structure to numerically solve the coupled electromechanical equations (see Appendix I) \cite{uetsuji2011multiscale}. For these calculations we use the measured value of permittivity as a fixed input ($\varepsilon_\text{eff} = 144.6$ at 30 V bias, see the following section). The resulting effective piezoelectric coefficients are then incorporated into the FEM acoustic model to evaluate the electromechanical coupling $k_{\text{eff}}^2$. 
The single-domain piezoelectric coefficients are extracted by fitting the Sezawa mode's measured angle-dependent $k_{\text{eff}}^2$, whose orientation response is much stronger than that of the Rayleigh mode,  as shown in Fig. \ref{fig:orientation}d. The best fit yields, with 95\% confidence interval (CI), $d_{31}$ = -47$\pm$5 pC/N, $d_{33}$ = 79$\pm$8 pC/N, and $d_{15}$ = 205$\pm$31 pC/N, of similar magnitude to those of  bulk BTO \cite{zgonik1994dielectric}. Simulations using these values reproduce the observed minimum in $k_{\text{eff}}^2$ near 45$^{\circ}$. From these, we obtain an effective piezoelectric constant $d_{33,\text{eff}}$ of 53$\pm$5 pC/N for the multi-domain BTO thin film at 0$^{\circ}$, roughly an order of magnitude larger than that of lithium niobate \cite{tian2024}. 

\section*{Ferroelectric switching and reconfigurability}

The ferroelectricity and relatively low Curie temperature of BTO also make its acoustic response reconfigurable in situ. Its low-voltage, switchable polarization, which distinguishes BTO from other piezoelectric materials such as AlN, AlScN, and LiNbO$_3$, is a tunable resource relevant both to classical switchable filters \cite{koohi2020reconfigurable} and, as discussed below, to tunable elements in cryogenic quantum circuits. We exploit this switchability to reconfigure the acoustic response. As illustrated in Fig. \ref{fig:switching}a, the initial balance of opposite domains at 0 V produces a negligible piezoelectric response. Increasing the applied voltage progressively switches more domains, leading to a gradual increase in the electromechanical coupling. Note that the periodic ripples superimposed on the resonance are due to BAW modes from the Si substrate (see Appendix A).

As the bias is swept up and down, the response is hysteretic, reflecting the lag in domain switching relative to the field \cite{acosta2017batio3}. Figure \ref{fig:switching}b-d capture this hysteresis through three observables of the same switching process: the IDT capacitance $C_0$, the mechanical resonance frequency, and the electromechanical coupling $k_{\text{eff}}^2$, each tracing the characteristic ferroelectric butterfly curve \cite{acosta2017batio3, geler2022ferroelectric}. These curves yield the film's main ferroelectric parameters. The bias at which the coupling vanishes gives a coercive field $E_c \approx 2.2$ MV/m, while a small residual coupling persists at 0 V from the remanent polarization $P_r$ \cite{zhao2019depolarization}. The saturated capacitance at 30 V corresponds to an effective relative permittivity $\varepsilon_\text{eff}$ of 144.6$\pm$2.6 (Appendix F), the value used to extract the piezoelectric coefficients above. The tuning itself is large: the resonance shifts by about 5\% (260 MHz) through electromechanical stiffening as the coupling turns on, and $k_{\text{eff}}^2$ swings from near zero to its fully-poled value, then decreases again as the bias is lowered owing to the depolarization field \cite{zhao2019depolarization}. Relative to switchable AlScN resonators \cite{mo2022complementary}, the two-orders-of-magnitude smaller coercive field \cite{eltes2020integrated} and roughly threefold larger $d_{33}$ of BTO allow in-situ switching while maintaining a high $k_{\text{eff}}^2$ (Appendix D).

A fast switching speed is important for real-time reconfiguration, particularly as the number of switching components increases \cite{koohi2020reconfigurable}. Domain switching in BTO films begins with the nucleation of domains with opposite polarization, followed by their growth through domain-wall motion \cite{geler2022ferroelectric}. The switching time is primarily limited by the nucleation time, which depends on several film properties, including strain, material defects, crystal orientation, and domain structure \cite{geler2022ferroelectric}. We measure the tuning dynamics in the time domain using the setup shown in Fig. \ref{fig:switching}e. The acoustic resonance is switched between two states by a square wave, with the transition captured by homodyne detection of the reflected RF signal magnitude. Figure \ref{fig:switching}f shows the dynamics when the bias voltage switches from zero to various positive biases $V_{\text{pp}}$. Note that the magnitude of the reflected signal decreases as the bias increases. The dynamics is well described empirically by a two-time-constant exponential, with a dominant fast component ($\tau_1\sim$60 ns) followed by a slower tail ($\tau_2\sim$1 $\mu$s) (see Appendix G). The switching time (defined as the 10-90\% transition) is 170 ns. Such fast, low-voltage tuning could support new functionalities across both classical and quantum domains. In the acoustic domain, spatiotemporal modulation could enable active stabilization and compact magnetless circulators \cite{estep2016magnetless}. More broadly, the voltage-tunable nonlinearity demonstrated here is of the kind recently proposed for cryogenic three-wave mixing and parametric amplification \cite{rosenthal2026paraelectric}, and the following section shows that this tunability persists down to millikelvin temperatures.

\begin{figure*}[ht]
    \centering
    \includegraphics[width=15.5cm]{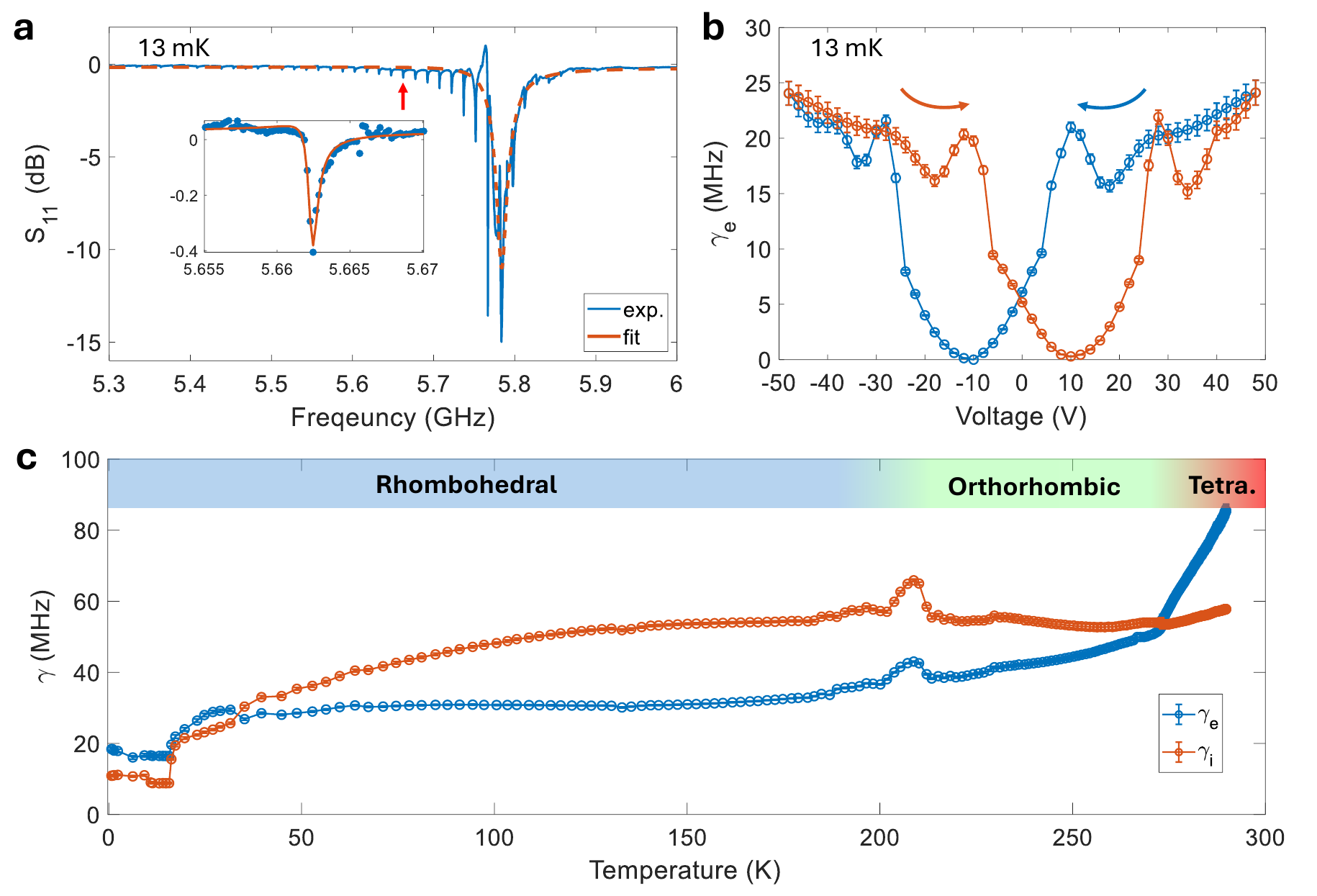}
    \caption{\textbf{Temperature dependence of BTO's piezoelectric response} \textbf{a,} Measured $S_{11}$ spectrum (blue) at 13 mK together with fitting (red dashed line) used to extract external coupling $\gamma_e$. Inset highlights one of the bulk acoustic modes (red arrow) with a fitted linewidth of 80 kHz. \textbf{b,} Hysteresis curve of $\gamma_e$ at 13 mK up to 50 V, showing an increased coercive field and remanent polarization. \textbf{c,} External ($\gamma_e$, blue) and internal ($\gamma_i$, red) coupling rates as a function of temperature from 300 K to 0.7 K. Shaded color band marks the corresponding phase transition of BTO. The error bars in \textbf{b} and \textbf{c} are 95\% CI from nonlinear fitting. The bias in \textbf{a} and \textbf{c} is 30 V.}
    \label{fig:temperature} 
\end{figure*}

\section*{Cryogenic characterization}
To assess the thin-film BTO for quantum transduction, we characterize its piezoelectric response at cryogenic temperatures. The IDT is cooled to 13 millikelvin (mK) in a dilution refrigerator (see Appendix J), and the $S_{11}$ spectrum is measured as shown in Fig. \ref{fig:temperature}a. The Sezawa mode exhibits a much narrower linewidth while remaining over-coupled, with $\gamma_e$ = 21 MHz and $\gamma_i$ = 12 MHz. Reduced phonon-phonon scattering in the Si substrate also reveals sharp BAW resonances with sub-MHz linewidth \cite{maccabe2020nano}. 
The hysteresis of the piezoelectric response at cryogenic temperature is shown in Fig. \ref{fig:temperature}b. A pronounced hysteresis in $\gamma_e$ is observed, with the coercive field elevated to 11 MV/m. The remanent polarization, reflected in the appreciable $\gamma_e$ at 0 V, also increases due to reduced thermal activation for domain nucleation and wall motion \cite{geler2022ferroelectric}.
%The bumps on the hysteresis curve are likely due to the interference of a spurious mode from the microwave setup which has a close frequency (see supplementary note xx).

The temperature dependence of $\gamma_i$ and $\gamma_e$ is illustrated in Fig. \ref{fig:temperature}c. The kink in $\gamma_e$ at 273 K aligns with the known tetragonal-orthorhombic phase transition \cite{eltes2020integrated, acosta2017batio3}, while the enhancement in $\gamma_e$ around 209 K may be associated with an increased piezoelectric response near the orthorhombic-rhombohedral phase transition \cite{anderson2025quantum}. The intrinsic loss $\gamma_i$ decreases by a factor of six at cryogenic temperature, due to the suppression of temperature-dependent mechanisms such as phonon-phonon scattering \cite{maccabe2020nano} and thermoelastic damping \cite{lifshitz2000thermoelastic}. Further investigation of cryogenic acoustic loss mechanisms in thin-film BTO would be valuable for quantum applications. 

Since direct extraction of admittance and $k^2_\text{eff}$ requires complex calibration apparatus that is not compatible with our current cryogenic setup (Appendix J), we estimate the cryogenic piezoelectric coefficient from the scaling of $\gamma_e \propto d_{\mathrm{eff}}^2/(s_{\mathrm{eff}}\varepsilon_{\text{eff}})$ \cite{blesin2021quantum}. Here $\gamma_e$ is measured directly and the compliance $s_{\mathrm{eff}}$ is essentially unchanged (the resonant frequency shifts by less than 3\%), so the inferred coefficient depends only weakly on the permittivity, as $d_{\mathrm{eff}}\propto\sqrt{\gamma_e\varepsilon_{\text{eff}}}$. We adopt the permittivity reduction (1.7$\times$) reported for BTO at 4 K, which is larger at lower biasing \cite{eltes2020integrated}. Given our higher poling field, it likely overestimates the actual reduction (Appendix J). Under this conservative input, the measured 4.6-fold drop in $\gamma_e$ gives a millikelvin $d_{33,\text{eff}}$ of about 19 pC/N, rising to $\sim$25 pC/N in the limit of unchanged permittivity. We therefore report $\sim$19 pC/N as a conservative estimate, a substantial fraction of the room-temperature value and consistent with the low-temperature decrease of the electro-optic coefficient measured in the same material \cite{eltes2020integrated}. More directly, the large measured $\gamma_e$ shows that the electromechanical coupling itself, not only the inferred coefficient, remains strong at millikelvin temperatures, which is ultimately what matters for piezoelectric coupling to quantum circuits.

\begin{table*}[t]
\centering
\begin{threeparttable}
\caption{Comparison of figure of merit $K^2$ between LiNbO$_3$, strontium titanate (STO), and BTO. For STO and BTO thin films which are multi-domain, the effective $d_{33,\text{eff}}$ and $s_{33,\text{eff}}$ are used. }
\begin{tabular}{p{1.9cm} p{1.5cm} p{1.5cm} p{1.5cm} p{1.5cm} p{1.5cm} p{1.5cm} p{1.5cm}}
\toprule
Material  & LiNbO$_3$ & STO (film) & STO (bulk) &  \multicolumn{2}{l}{BTO (film)} & \multicolumn{2}{l}{BTO (single domain)} \\ \hline
Ref.  & \cite{warner1967determination} & \cite{khalil2026cryogenic} & \cite{anderson2025quantum} &  \multicolumn{2}{l}{This work}  & \multicolumn{2}{l}{This work} \\ 
Temp. & RT/Cryo. & Cryo. & Cryo. & RT & Cryo. & RT & Cryo.  \\ 
$d_{33}$ (pm/V) & 6 & 27 & 91 & 53 & 19 & 79 & 28  \\ 
$s_{33}$ (1/Pa) & 5$\times10^{-12}$ & 3.8$\times10^{-12}$ & - & 10$\times10^{-12}$ & 10$\times10^{-12}$ & 13$\times10^{-12}$ & 13$\times10^{-12}$   \\ 
$\varepsilon_{r,33}$ & 30 & 1200 & 7500 & 145 & 85\tnote{b} & 129\tnote{a} & 76\tnote{b}  \\
$K^2$ & 0.027 & 0.018 & 0.023 & 0.219 & 0.048 & 0.421 & 0.09 \\
\bottomrule
\end{tabular}
\label{table_1}
\begin{tablenotes}
\footnotesize
\item[a] Adapted from Ref. \cite{zgonik1994dielectric}.
\item[b] Estimation based on the scaling of permittivity at cryogenic temperature measured in Ref. \cite{eltes2020integrated}.
\end{tablenotes}
\end{threeparttable}
\end{table*}

\section*{Discussions and outlook}

In summary, we have used a simple IDT SAW platform to characterize the mechanical and piezoelectric properties of thin-film BTO from room temperature down to millikelvin. At room temperature, the resonators reach an electromechanical coupling $k_{\text{eff}}^2$ of 14.2\% at 5.2 GHz and operate up to 7.8 GHz, with a low-voltage reconfigurability suited to switchable RF filters. The extracted piezoelectric coefficients are comparable to those of bulk BTO and persist down to millikelvin temperatures, with a two- to threefold reduction, where the material remains promising for quantum acoustics. These results position thin-film BTO as a piezoelectric platform spanning the full temperature range relevant to classical RF and cryogenic quantum hardware.

% To place BTO among other piezoelectric materials, we compare through the figure of merit $K^2$ rather than any single coefficient, since $K^2$ governs the achievable electromechanical coupling and absorbs the large permittivity differences across these materials (Table \ref{table_1}) \cite{blesin2021quantum}. We base the comparison on $d_{33,\text{eff}}$ because our cryogenic measurement determines its corresponding coupling directly. We report the effective, multi-domain value relevant to real devices, with single-domain and bulk values listed in Table \ref{table_1} for reference. A comparison through other channels, such as the shear coupling (governed by $d_{15}$), would require compliance and permittivity components we do not measure here; because BTO, LiNbO$_3$, and STO all have $d_{15}$ substantially larger than $d_{33}$ while the permittivities entering the shear coupling differ from material to material, the outcome of such a comparison is not obvious a priori. Restricting to the $d_{33}$-mediated coupling that our devices and cryogenic measurement access, the interplay between a moderate piezoelectric response and a relatively low permittivity lets BTO compare favorably even with materials that have larger $d_{33}$, such as strontium titanate \cite{anderson2025quantum, khalil2026cryogenic}.

To place BTO among other piezoelectric materials, we compare through the figure of merit $K^2=d^2/(s\varepsilon_r)$ rather than any single coefficient, since $K^2$ governs the achievable electromechanical coupling and absorbs the large permittivity differences across these materials (Table \ref{table_1}) \cite{blesin2021quantum}. We base the comparison on $d_{33,\text{eff}}$ because our cryogenic measurement determines its corresponding coupling directly \footnote{A comparison through other channels, such as the shear coupling governed by $d_{15}$, would require compliance and permittivity components we do not measure here. Because BTO, LiNbO$_3$, and STO all have $d_{15}$ substantially larger than $d_{33}$, while the permittivities entering the shear coupling differ from material to material, the outcome of such a comparison is not obvious a priori.}. We report the effective, multi-domain value relevant to real devices, with single-domain and bulk values listed in Table \ref{table_1} for reference. Restricting to the $d_{33}$-mediated coupling, the interplay between a moderate piezoelectric response and a relatively low permittivity lets BTO compare favorably even with materials that have larger $d_{33}$, such as strontium titanate \cite{anderson2025quantum, khalil2026cryogenic}.

Several directions could push the platform further. The most direct path to stronger coupling is single-domain BTO, recently demonstrated by lattice engineering \cite{lee2021plane} or by spalling from bulk \cite{thureja2025spalled}, which would move the response toward the single-domain values in Table \ref{table_1}. Mechanical loss is equally important for quantum applications, and in our devices it is set by radiation into the substrate rather than by intrinsic material loss. It can be reduced with higher acoustic-contrast substrates such as diamond or SiC \cite{zhang2020surface}, or removed in suspended resonators \cite{tian2024, anderson2026high}, which would additionally allow the intrinsic mechanical loss of BTO to be measured directly at millikelvin temperatures \cite{emser2024thin}.

Beyond improving these devices, the same material gives access to additional modes and applications. Alternative poling schemes and crystal cuts, such as c-domain films, would reach the fundamental SAW mode with higher mechanical $Q$, shear modes that exploit the larger $d_{15}$ \cite{nordlander2020, zgonik1994dielectric}, and bulk modes for millimeter-wave operation \cite{xie2025towards}. Integrated devices, including nanoscale transducers for nano-electromechanical systems \cite{craighead2000nanoelectromechanical} and microwave-to-optical conversion \cite{mirhosseini2020superconducting}, together with low-loss acousto-optic devices that exploit the large photoelastic response of BTO \cite{zgonik1994dielectric, tian2024}, will require further progress in high-quality BTO etching, for which recent argon-ion-milling results are encouraging \cite{raju2025high}. With these advances, thin-film BTO could become a shared platform for reconfigurable RF, acousto-optic, and cryogenic quantum devices.

\section*{Methods}

\noindent \textbf{BVD model} The Butterworth-Van Dyke (BVD)
model has been widely used to model the electrical response for piezoelectric resonators \cite{blesin2021quantum}. As shown in Fig. \ref{fig1}d, it consists of an RLC resonator to model the mechanical resonant, in parallel with a capacitor $C_0$ modeling the intrinsic capacitance from the IDT electrodes. Multiple mechanical resonances can be modeled by adding extra RLC branches for each resonance. 
A good matching between the BVD model fitting and the experimental admittance can be obtained (see Fig. \ref{fig1}d). Because of $C_0$, the admittance for each mechanical resonance shows the resonance ($f_s$) and anti-resonance ($f_p$). 
The $k_{\text{eff}}^2$ is conventionally calculated from the difference between $f_s$ and $f_p$, which is also related to the ratio between the motional capacitance $C_m$ and $C_0$ \cite{dahmani2020piezoelectric}:
\begin{equation}
    k_{\text{eff}}^2=\frac{\pi^2}{8}\left(\frac{f^2_p-f^2_s}{f^2_s}\right)=\frac{\pi^2}{8}\frac{C_m}{C_0}
\end{equation}
Therefore, it can be seen the larger the separation between $f_s$ and $f_p$ and larger ${C_m}$, the higher the $k_{\text{eff}}^2$, which is important for designing acoustic filters with higher bandwidth \cite{koohi2020reconfigurable}. 
\\
\\
\noindent \textbf{Data availability} 
The datasets generated during and/or analyzed during the current study are available from the corresponding author (H.T.) upon reasonable request.
 \begin{acknowledgments}
We acknowledge A. Bozkurt for valuable discussions, and F. Yang,
and P. Shah for help in cryogenic setup. This work was supported by the NSF (Award Nos. 2137776 and 2238058, QuIC). H.T. gratefully acknowledges support from the IQIM Postdoctoral Fellowship. S.Y.C gratefully acknowledges support from the Gupta fellowship. We gratefully acknowledge the critical support and infrastructure provided for this work by The Kavli Nanoscience Institute at Caltech. 
\\
\\
\textbf{Author contributions} 
M.M., H.T., and S.Y.C. conceived and designed the experiment.
H.T., S.Y.C., and N.A. performed the numerical simulations of the devices. H.T. fabricated the devices. H.T. and S.Y.C. conducted room temperature measurement and analysis. S.Y.C. built the multi-domain mechanical and piezoelectric models. S.Y.C. and H.T. conducted cryogenic measurement and data analysis. H.T., S.Y.C., and M.M. wrote the paper with input from all authors. K.S. contributed to thin-film characterization methods and the understanding of thin-film stress and domain structure that informed the device design. M.M. supervised the project.
\\
\\

\end{acknowledgments}
\clearpage

%% supplementary
\appendix
\onecolumngrid
\renewcommand{\thefigure}{S\arabic{figure}}
\setcounter{figure}{0}
\renewcommand{\thetable}{S\Roman{table}}
\setcounter{table}{0}

\section{Bulk acoustic modes from Si substrate}

\begin{figure*}[h]
    \centering
    \includegraphics[width=15.5cm]{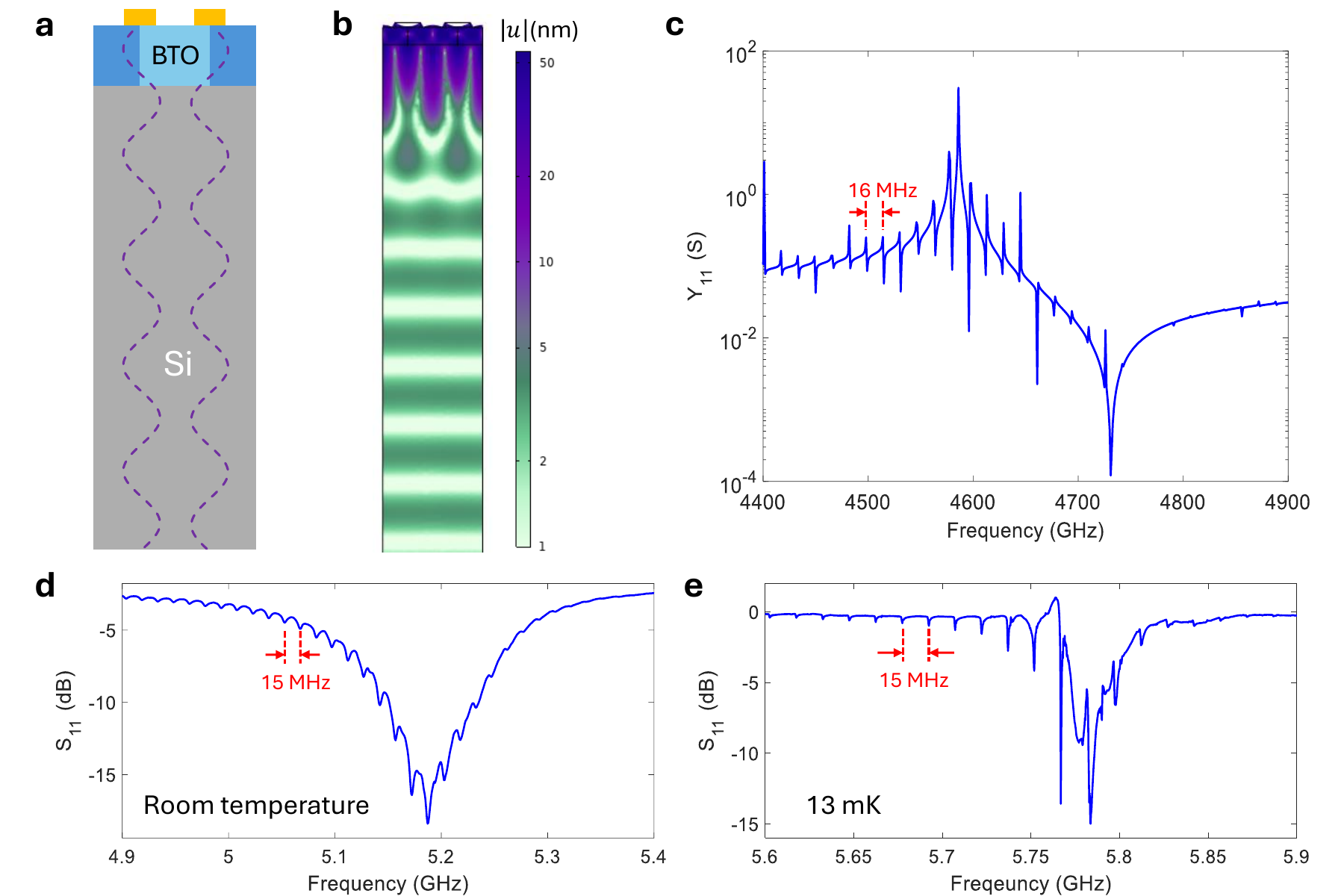}
    \caption{\textbf{Bulk acoustic waves from the Si substrate} \textbf{a,} Schematic of the cross-section of one period of IDT. BAW (purple dashed curve) are resonant in the Fabry-Pérot cavity formed by the top and bottom surfaces of the substrate. \textbf{b,} Numerical simulation of the Sezawa mode profile. The color bar is the magnitude of the displacement in logarithm scale, where the BAW in the substrate can be clearly seen. Note only the zoom-in near the top few microns of the substrate is shown. \textbf{c,} Simulated admittance $Y_{11}$ of the Sezawa mode, where resonances of BAW are found with period of 16 MHz. \textbf{d,} Measured $S_{11}$ of the Sezawa mode at \textbf{d,} room and \textbf{e,} cryogenic temperatures, respectively. FSR of 15 MHz is obtained for the BAW modes.  }
    \label{Fig:bulk mode} 
\end{figure*}

As mentioned in the main text, the acoustic loss of the SAW is dominated by radiation into the bulk substrate. In this section, the existence of BAW modes are verified through numerical simulation, which shows good agreement with the experiment. As shown in Fig. \ref{Fig:bulk mode}a, the top and bottom surfaces of the substrate naturally form a Fabry-Pérot cavity which supports resonances that confine the acoustic waves radiated into the substrate. These acoustic modes are conventionally referred to as high-overtone bulk acoustic resonances (HBAR) \cite{tian2024, blesin2021quantum, luo2025lifetime}. In the simulations of the main text, the Si substrate is assumed to be infinite with low reflecting boundary condition at the Si bottom surface. Therefore, the HBARs are not found. By including Si substrate with the actual thickness as the sample (280 $\mu$m) in the simulation, BAWs are clearly seen in Fig. \ref{Fig:bulk mode}b. The simulation is done at the resonance of Sezawa mode. The spectrum of the admittance is calculated in Fig. \ref{Fig:bulk mode}c, where, on top of the resonance and anti-resonance of the Sezawa mode, there is a series of HBAR with period of 16 MHz. This is in good agreement with the free spectrum range (FSR $\simeq$15 MHz) observed in the $S_{11}$ spectra at room (Fig. \ref{Fig:bulk mode}d) and cryogenic (Fig. \ref{Fig:bulk mode}e) temperatures. It can be seen the HBARs become prominent at 13 mK as the phonon dissipation in Si substrate decreases at low temperature, especially for the ones that are close to the Sezawa mode. This poses difficulty in fitting for the Lorentzian of the Sezawa mode with high quality. These HBAR modes can be suppressed by adopting substrates with higher acoustic velocity such as diamond and SiC for larger acoustic contrast between BTO \cite{zhang2020surface}.

\section{Device and chip variations}
\begin{figure*}[t]
    \centering
    \includegraphics[width=15.5 cm]{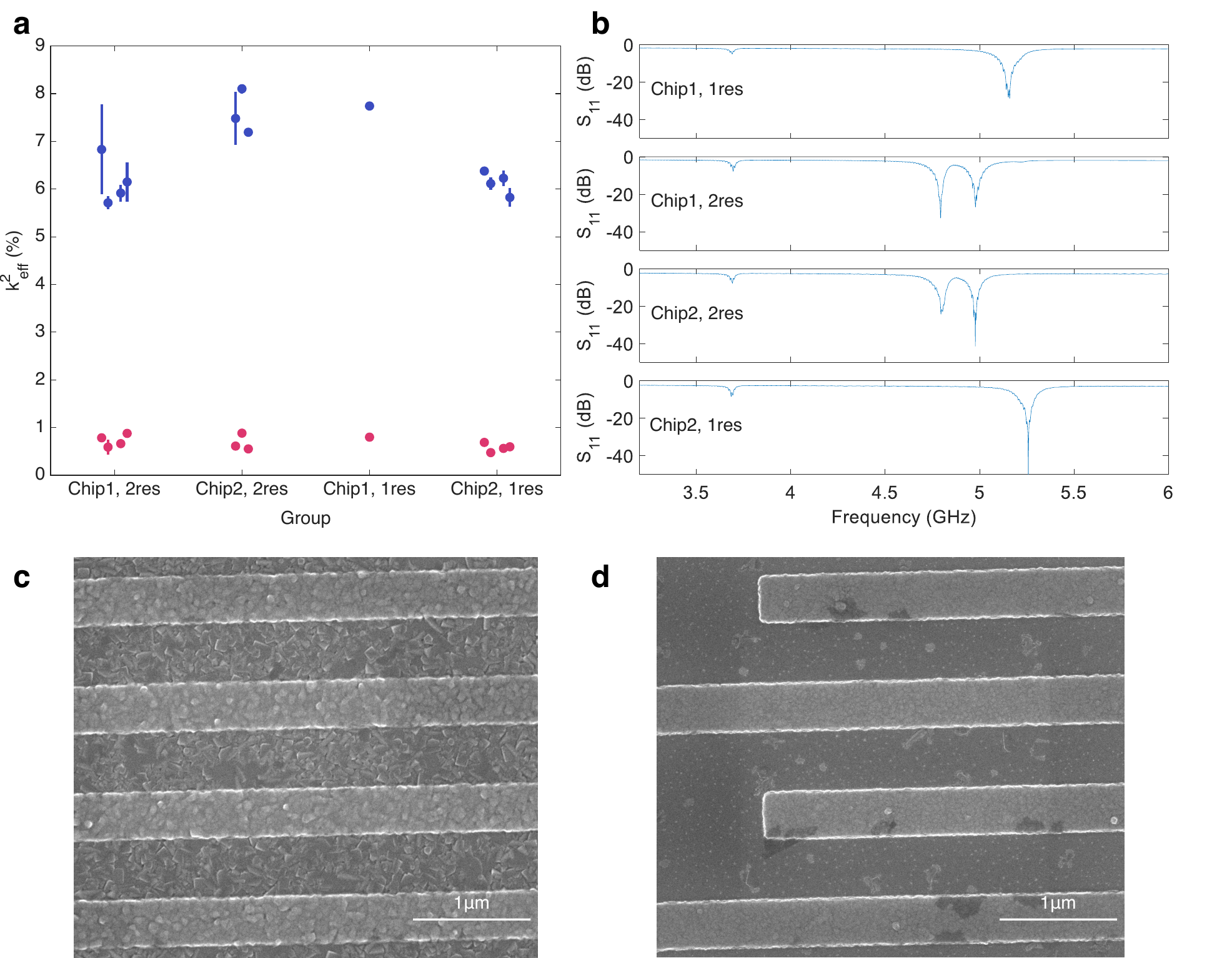}
    \caption{\textbf{Performance variation among devices and chips.}  \textbf{a,} $k_{\text{eff}}^2$ measurements of multiple devices with the same design across different locations on two chips. Across both chips, there are two groups of devices with distinct mechanical responses, as shown in \textbf{b,}. One group of devices exhibit strong coupling between the shear and Sezawa modes as they have close frequencies (referred to 2 res.), while the other group shows a much lower shear mode frequency (referred to 1 res.). Each dot represents each device. Error bars are the standard deviation of three repeated measurements. SEM images of the surface for 1-resonance \textbf{c,} and 2-resonance \textbf{d,} devices on chip 2, showing clear contrast of the surface texture between the two types of devices. }
    \label{Consistency} 
\end{figure*}

Spatial non-uniformity of the BTO film such as surface texture and crystal disorder can modify the local domain microstructure, consequently effecting the electromechanical response of the IDT. To assess the reproducibility of the IDT response and the measured $k^2_{\text{eff}}$, we characterize the $S_{11}$ response of devices with nominally identical design ($0^{\circ}$ orientation, 40 pair and 30 $\mu$m IDT) at multiple locations on two separately fabricated chips (hereafter referred to as Chip 1 / Chip 2). Within each chip, two types of mechanical responses are observed, as shown in Fig. \ref{Consistency}b. One group of devices shows similar response as the device presented in the main text, where the shear and Sezawa modes are uncoupled and far away from each other. We also observe devices showing two close resonances for the shear and Sezawa modes. The hybridization with the Sezawa mode enhances the electromechanical coupling of the shear mode. 

The $k^2_{\text{eff}}$ of Sezawa and Rayleigh modes for these two types of devices across the two chips is summarized in Fig.\ref{Consistency}a. It can be seen the $k^2_{\text{eff}}$ of Sezawa varies between 6 and 8\%. In the main text, we primarily present the device with a clear separation between shear and Sezawa modes to avoid ambiguity in identifying the two modes due to mode hybridization. 

%We note that the difference between these two types of devices is mainly the shear mode's frequency. 
As shown in Fig. \ref{Consistency}c-d, SEM images show distinct surface texture of these two devices, with one having smooth surface while the other showing dense surface roughness \cite{posadas2021thick}. This non-uniformity of crystalline structure of the BTO film could potentially explain the appearance of devices showing distinct mechanical resonances. Further numerical simulations reveal that the variation of the effective shear compliance $s_{44}$ can explain the changes of shear frequency, which is found to be related with the presence of c-domains (see supplementary note I). 

\begin{figure*}[ht]
    \centering
    \includegraphics[width=14 cm]{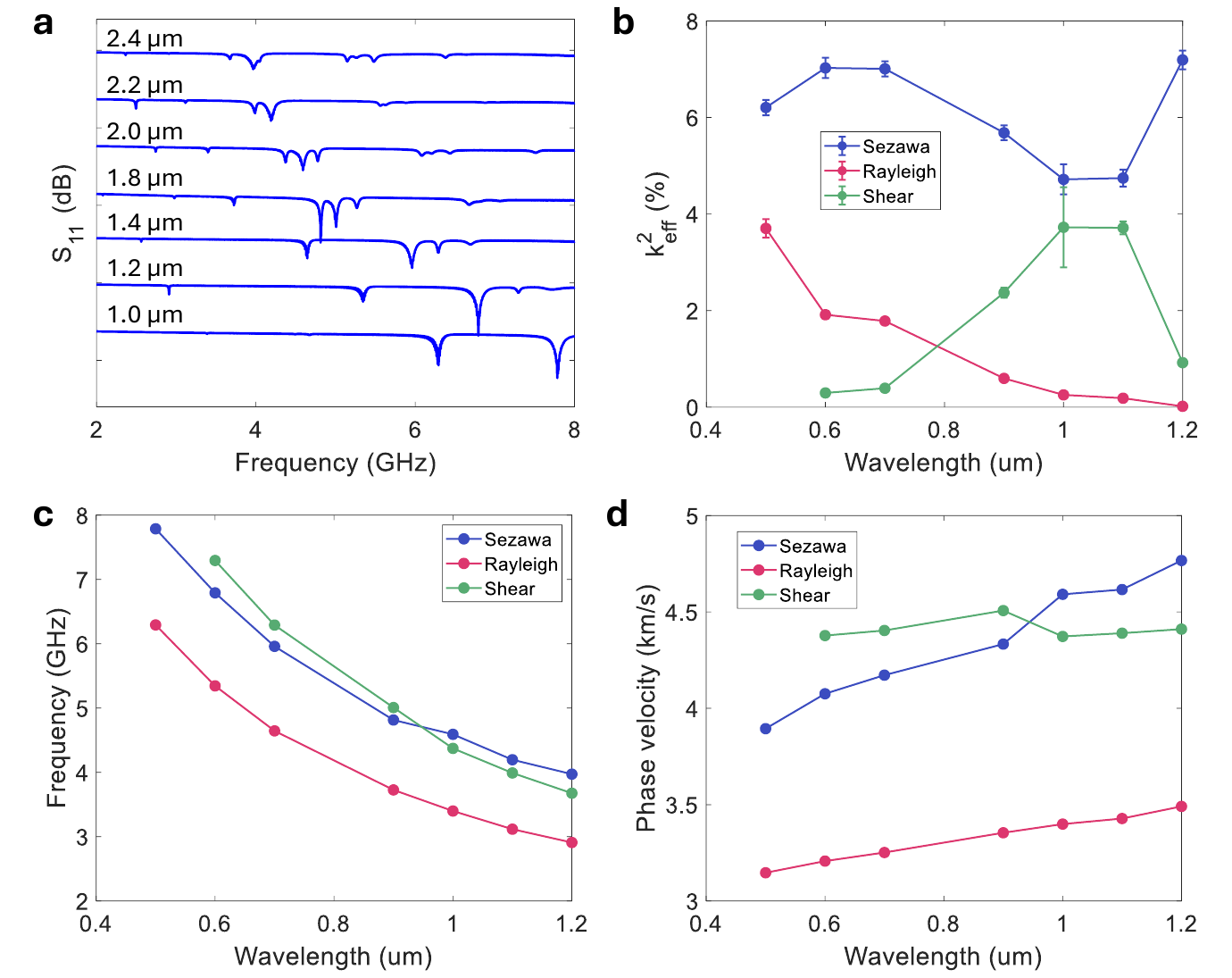}
    \caption{\textbf{Frequency scaling of the BTO SAW with IDT period.} \textbf{a,} $S_{11}$ spectra for IDTs with various IDT pitches as labeled. Each trace is vertically shifted for better illustration. The spacing between y-ticks is 50 dB. The bias voltage is 30 V for all devices. \textbf{b,} $k^2_{\text{eff}}$ as a function of acoustic wavelength for Sezawa, Rayleigh, and shear modes. Error bars are the standard deviation of three repeated measurements. \textbf{c,} Resonant frequency and \textbf{d,} phase velocity for BTO SAW with various wavelength. The mode anti-crossing between Sezawa and shear modes can be seen. }
    \label{period} 
\end{figure*}

\section{Dependence on IDT period, pair number and aperture length}

\subsection{IDT period and frequency scaling}
High frequency operation while still maintaining high electromechanical coupling is essential for large bandwidth acoustic filters in next-generation wireless communication systems \cite{koohi2020reconfigurable}. Here, the frequency scaling of the BTO SAW is studied by characterizing IDTs with various periods. Figure \ref{period}a shows the $S_{11}$ spectra for IDT period ranging from 1 to 2.4 $\mu$m, corresponding to wavelength from 0.5 to 1.2 $\mu$m. The acoustic frequency increases with the decreasing in wavelength, as shown in Fig. \ref{period}c for Sezawa, Rayleigh, and shear modes. It is noted that there is mode anti-crossing between Sezawa and shear modes when they hybridize around wavelength of 1 $\mu$m. The mode anti-crossing can be seen more clearly in the plot of phase velocity for the three modes in Fig. \ref{period}d. The velocities for Sezawa and Rayleigh modes increase with wavelength, since the participation of Si increases for larger acoustic modes at longer wavelength. 

The $k^2_{\text{eff}}$ at each wavelength is calculated as shown in Fig. \ref{period}b.
Each device is measured three times such that the error bar includes variations due to the RF probe contact and nonlinear fitting using the BVD model. Due to the mode hybridization, there is a drop in $k^2_{\text{eff}}$ for Sezawa mode while an enhancement for shear mode. This is behavior is verified through FEM simulations (not shown here). In the main text, we choose to study devices with large separation between Sezawa and shear modes to avoid mode hybridization. At 0.5 $\mu$m wavelength, high frequency Sezawa mode at 7.8 GHz is observed with a decent $k^2_{\text{eff}}$ of 6.2$\pm$0.2\%. It is also observed that the $k^2_{\text{eff}}$ of Rayleigh mode increases at shorter wavelength because of better mode overlap with BTO thin film. Higher $k^2_{\text{eff}}$ and frequency Rayleigh mode can be achieved by further decreasing the wavelength, which is ultimately limited by the resolution in lithography.

\begin{figure*}[ht]
    \centering
    \includegraphics[width=14 cm]{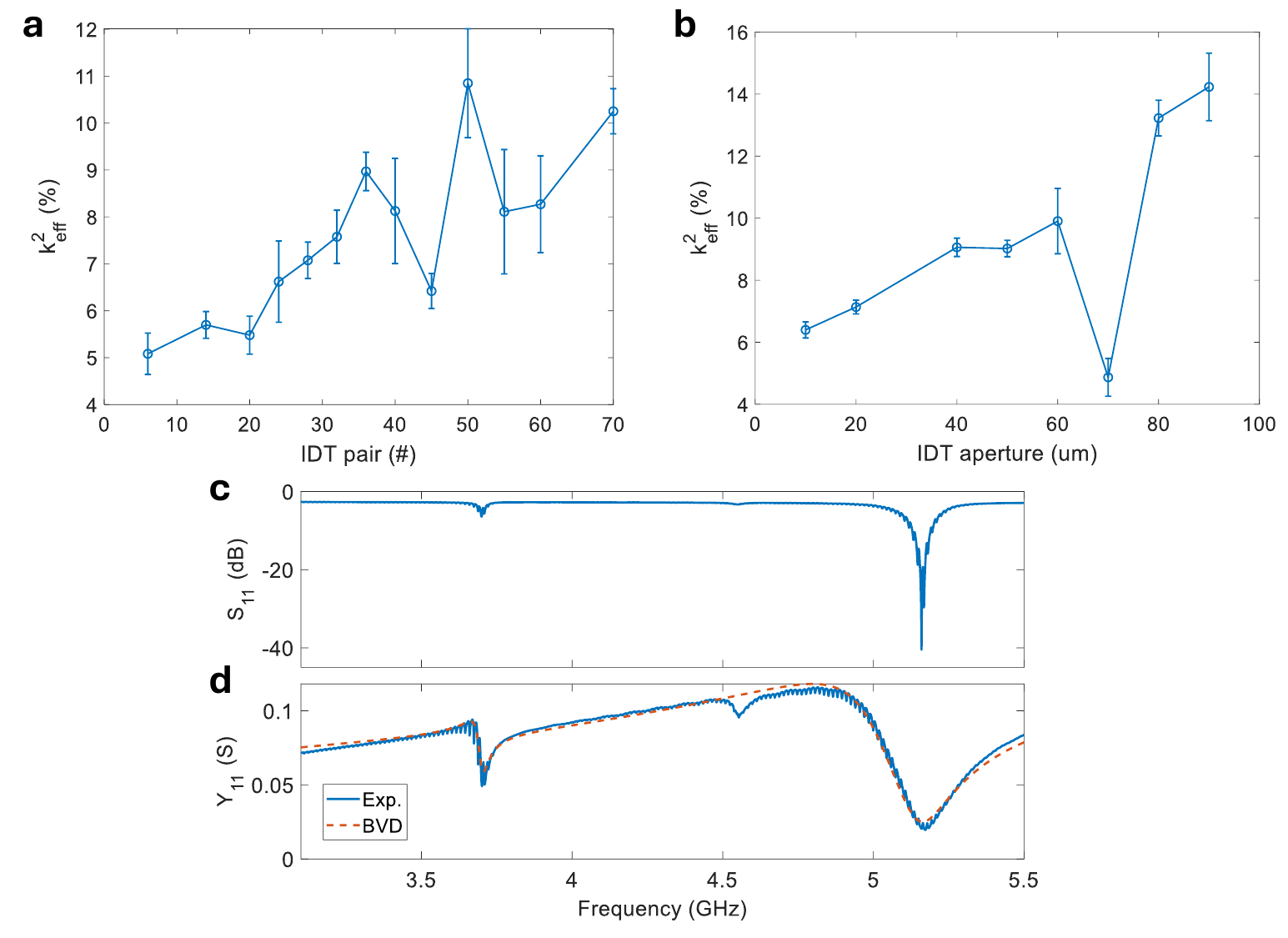}
    \caption{\textbf{Dependence of electromechanical coupling on IDT pair and aperture length.} \textbf{a,} Measured $k^2_{\text{eff}}$ for IDTs with different number of pairs while aperture is kept as 30 $\mu$m. \textbf{b,} Measured $k^2_{\text{eff}}$ for IDTs with different aperture lengths with 40 pairs. Error bars are the standard deviation of three repeated measurements. \textbf{c,} Experimental $S_{11}$ spectrum and \textbf{d,} $Y_{11}$ together with BVD model fitting (red dashed line) for IDT with 90 $\mu$m aperture and 40 pairs. $k^2_{\text{eff}}$ of 14.2$\pm$1.1\% is obtained for the Sezawa mode at 5.16 GHz. }
    \label{Length} 
\end{figure*}

\subsection{IDT pair and aperture length}
The performances of IDTs with different number of pairs and aperture lengths are studied to examine the influence on electromechanical coupling. Figure \ref{Length}a shows the dependence of $k^2_{\text{eff}}$ on IDT pair. It can be seen $k^2_{\text{eff}}$ increases and saturates around 10\% at large number of pairs. The $k^2_{\text{eff}}$ increases with IDT aperture as explored in Fig. \ref{Length}b. The dip at 70 $\mu$m is caused by the coupling of Sezawa mode to a spurious mode, which is less prominent in other devices.
Maximal $k^2_{\text{eff}}$ of 14\% is demonstrated with 90 $\mu$m aperture IDT. The experimentally measured $S_{11}$ and $Y_{11}$ responses of this device are presented in Fig. \ref{Length}c and d, respectively. Critical coupling with 40 dB $S_{11}$ depth is obtained for the Sezawa mode. 

It is also noted that as the pair and aperture increase, the intrinsic capacitance $C_0$ of the IDT increases, which elevates the background in $Y_{11}$ response and makes the resonance peaks less sharp and discernible (comparing Fig. \ref{Length}d and Fig. \ref{fig1}d in the main text). This poses challenge in BVD fitting with high fidelity. Therefore, in the main text we still use 40 $\mu$m aperture.

\section{Comparison with state-of-the-art acoustic resonators}
\begin{figure*}[ht]
    \centering
    \includegraphics[width=11 cm]{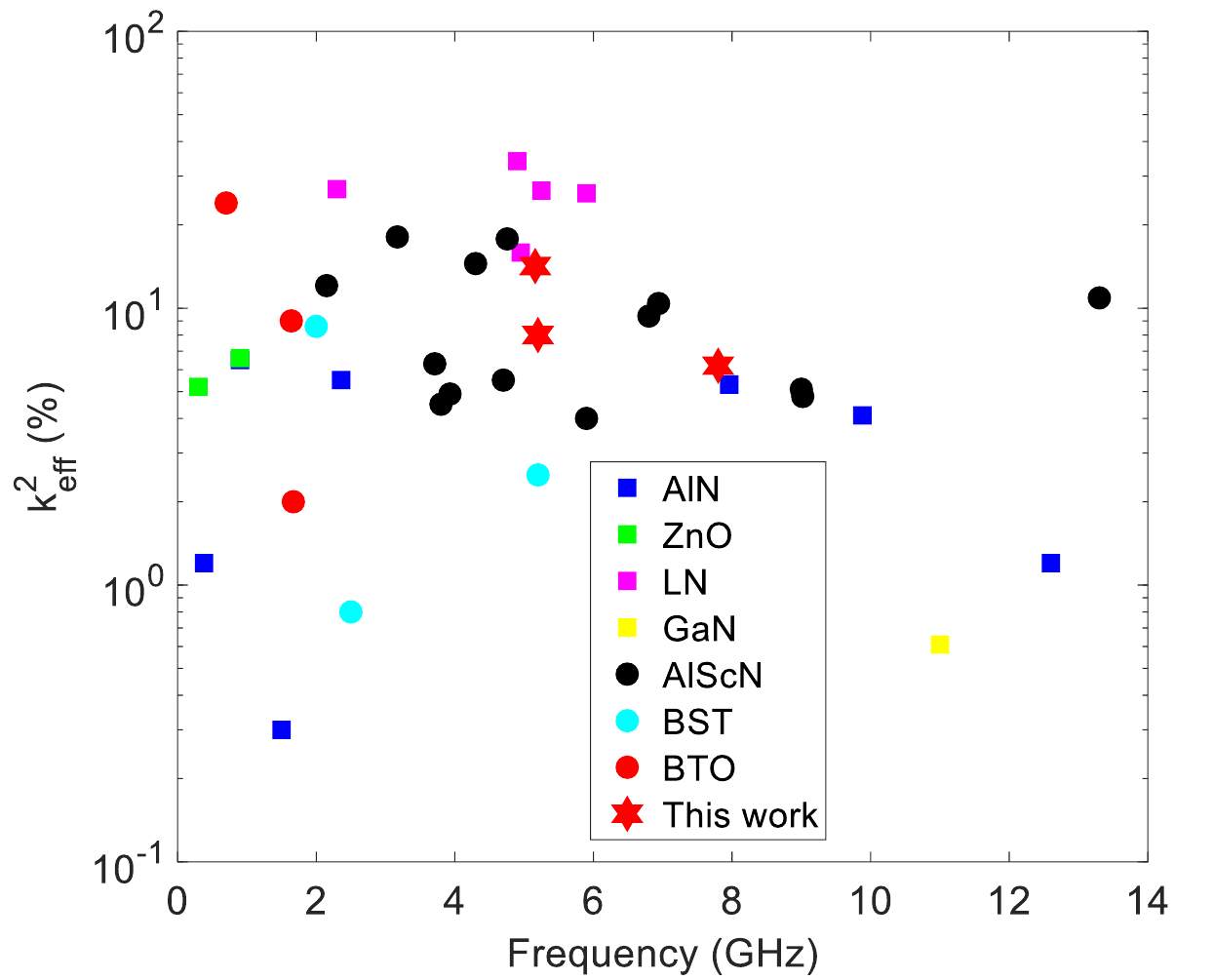}
    \caption{\textbf{Comparison with acoustic resonators made from different materials.} Different color represents different material as labeled. Square symbols are non-switchable resonators, while circle symbols are switchable. Our work is represented by red stars. The data is extracted from references \cite{du2024near, koohi2020reconfigurable, ahmed2023super, zhang2020surface, hadj2019sezawa, wang2020film, zou2022aluminum, moreira2011aluminum, lee2013intrinsically, rodriguez2012super, park201910, park2020epitaxial, dou2023super, lu2025switchable, nam2023mm, mo2022complementary, anderson2026high}}
    \label{comparison} 
\end{figure*}

The performance of our BTO SAW resonator is compared with acoustic resonators made from other materials as shown in Fig. \ref{comparison}. The highest $k^2_{\text{eff}}$ is achieved with SAW resonators made from LN with frequency as high as 6 GHz \cite{du2024near, dai2024coupled}. Scandium doped aluminum nitride (AlScN) has attracted much attention recently because of its improved ($\sim$5 times) piezoelectric coefficient compared with traditional AlN \cite{zou2022aluminum}. High $k^2_{\text{eff}}$ has thus been demonstrated using film bulk acoustic resonator (FBAR) \cite{wang2020film}. Although AlScN is ferroelectric, switching of AlScN resonator requires much higher voltage from 80 to 350 V \cite{wang2020film, mo2022complementary}, which poses challenges for chip level integration and compatibility with modern RF front ends. The two orders of magnitude smaller coercive field on BTO allows the demonstration of a low-voltage ($\leq$15 V) switchable BTO SAW resonator while maintaining $k^2_{\text{eff}}$ higher than AlScN SAW with similar design. It is noted that the switching voltage is dependent on the spacing between electrodes. For SAW, the spacing is limited by lithography resolution which is around a few hundreds of nanometer. The narrowest spacing of 500 nm is demonstrated in Fig. \ref{period}, which results in switching voltage smaller than 10 V. In the future, by utilizing BAW resonator where the ferroelectric film is sandwiched vertically between electrodes, the switching voltage can be largely reduced, since the electrode spacing is determined by the film's thickness which can be controlled through deposition. 

Compared with other low-voltage switchable resonators, such as BTO and barium strontium titanate (BST) resonators \cite{lee2013intrinsically, koohi2020reconfigurable, anderson2026high}, our work shows higher $k^2_{\text{eff}}$ and frequency, which could potentially meet the requirements in future reconfigurable RF systems (red stars in Fig. \ref{comparison}). Additionally, the SAW design in this work is targeted for the ease of material characterization. With optimized and more complicated device design and fabrication (such as released Lamb resonator \cite{anderson2026high}), higher $k^2_{\text{eff}}$ and mechanical-Q BTO resonators can be achieved in the near future.

\section{Leakage current at room and cryogenic temperatures}
\begin{figure*}[hb]
    \centering
    \includegraphics[width=14 cm]{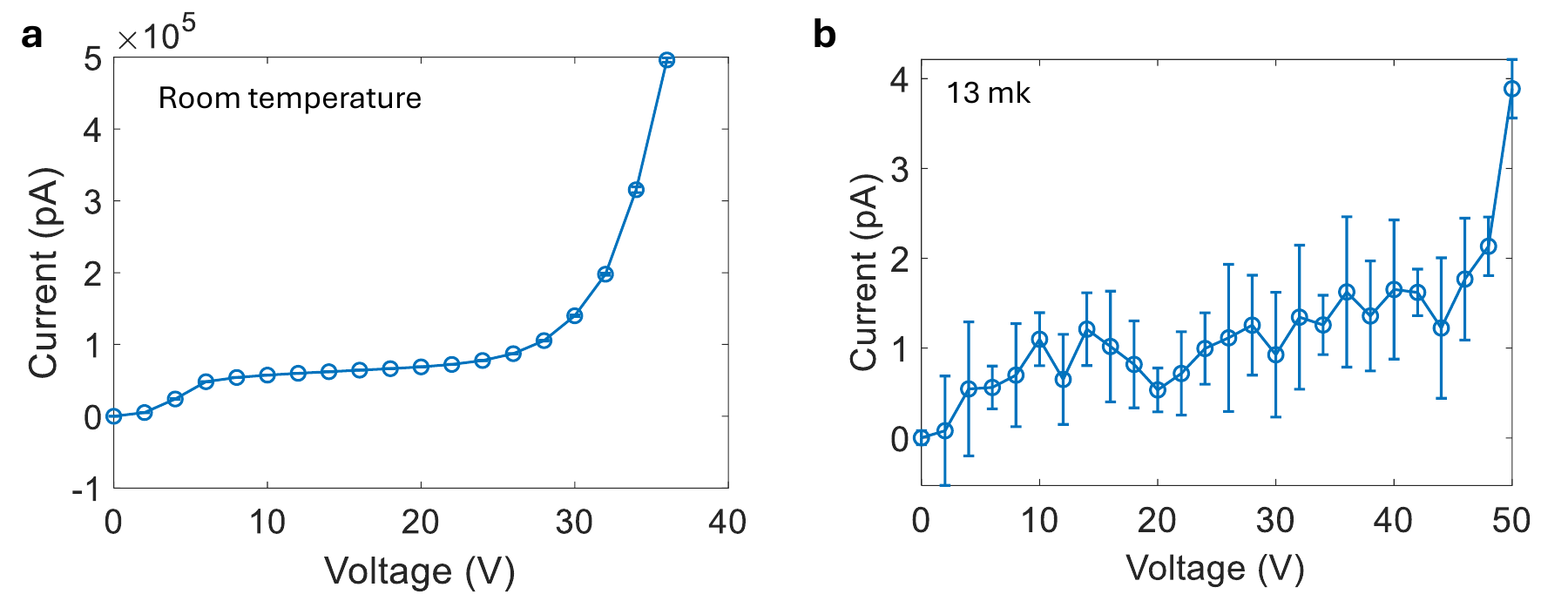}
    \caption{\textbf{Leakage current of the IDT.} Measured leakage current at \textbf{a,} room and \textbf{b,} cryogenic temperatures. The error bar is the standard deviation from three repeated measurements.  }
    \label{Fig:leakage} 
\end{figure*}

The static leakage current of the IDT with increasing bias voltages is shown in Fig. \ref{Fig:leakage}. At room temperature, the current starts to increase exponentially at 30 V. Low leakage current that is an order of magnitude smaller is reported for MBE-grown BTO before \cite{eltes2020integrated}. A few potential causes for the higher current can be identified. One leakage path could be from the finite resistivity of the Si substrate at room temperature especially when the bias field is approaching the breakdown field of Si which is around 30-50 MV/m \cite{kim2017dielectric}. The microstructure and defects in BTO films can also contribute to the higher leakage current. For example, oxygen vacancy is commonly observed in sputtered BTO thin films \cite{tyunina2020oxygen}. The grain boundaries and domain walls can also accumulate free charges that form conduction paths \cite{guo2018influence}.  
Further study of the leakage current of RF-sputtered BTO thin film is needed to evaluate its performance at high bias field. 

At cryogenic temperature, the leakage current is suppressed by 5 orders of magnitude to pA level (see Fig. \ref{Fig:leakage}b), which could be due to the freezing out of Si free carrier and lower mobility of oxygen vacancy. This allows a much higher bias voltage (up to 50 V) to be applied which is limited by the maximum voltage allowed by the cryogenic setup. It is noted that the measured pA level leakage is at the noise background of the measurement apparatus, such as the leakage through coaxial cables. Thus, the measurement gives an upper bound of the leakage current through BTO thin films. Previous study has shown a sub-pA leakage current of BTO at 4 K \cite{eltes2020integrated}.

\section{Calibration of in-plane BTO permittivity}
\begin{figure*}[ht]
    \centering
    \includegraphics[width=14 cm]{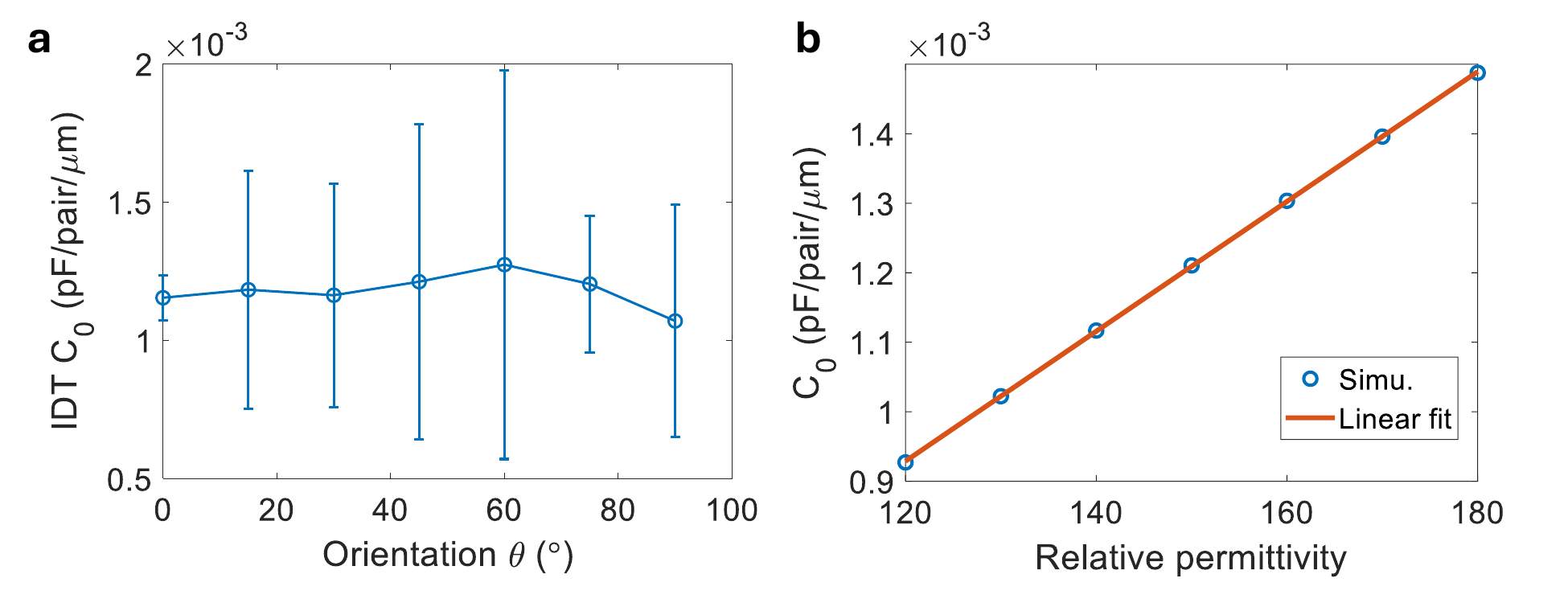}
    \caption{\textbf{IDT capacitance measurement and fitting} \textbf{a,} Experimentally measured intrinsic capacitance $C_0$ of IDT at different orientations. The error bar is the 95\% confidence interval of linear fitting of the admittance. Capacitance is measured at biasing of 25 V. \textbf{b,} Numerical simulation of $C_0$ by varying the in-plane relative permittivity.  }
    \label{Fig:C0} 
\end{figure*}

To better simulate the SAW response through numerical models, a good estimation of the relative permittivity $\varepsilon_r$ of BTO is essential to simulate the electric field. To experimentally measure the permittivity, the capacitance $C_0$ of IDT is obtained through calibrated admittance which is then compared with the $C_0$ simulated from the electrostatic modeling of one IDT electrodes pair \cite{abel2019large, chelladurai2025barium}. While $C_0$ can be extracted from the BVD model, we chose to linearly fit for admittance that is far from any mechanical resonances ($\sim$1 GHz) with $Y=j\omega C_0$, which generally gives better confidence interval. In the experiment, besides the IDT electrodes, there is parasitic capacitance from RF probe contact pads. To remove it, a series of IDTs with different number of pair is fabricated. Linear fitting of the $C_0$ with respect to pair number gives normalized $C_0$ with unit IDT pair and length, which can be directly simulated. 

The orientation dependence of $C_0$ is investigated as shown in Fig. \ref{Fig:C0}a. As there is no clear relation between $C_0$ and orientation, we take the average $C_0$ of 1.16$\pm$0.025 fF/pair/$\mu$m. Figure \ref{Fig:C0}b shows the simulated $C_0$ as function of in-plane $\varepsilon_r$, from where the experimental BTO $\varepsilon_r$ can be interpolated to be 144.6$\pm$2.6. As illustrated in Fig. 2b in the main text, the permittivity is related with the polarization state which is dependent on the bias voltage. The above measurement is done at 25 V. The permittivity at other biases can be scaled from the hysteresis of $C_0$. It is noted that the extracted $\varepsilon_r$ is on par with that obtained for MBE BTO film at (equivalently) 18 V, which is 198 \cite{eltes2020integrated}. 

As the electric field is primarily in the plane, the electrical and piezoelectric response is dominated by the in-plane permittivity. As the film is primarily a-domain, the out of plane permittivity is the a-axis permittivity of BTO $\varepsilon_{r,a}$. We take the number from the literature, which is about 2000 \cite{eltes2020integrated, chelladurai2025barium, zgonik1994dielectric}.  

\section{Domain switching dynamics}
\subsection{Measurement setup}
Our measurement setup is illustrated in Fig. \ref{fig:switching}e in the main text. A microwave signal at the Sezawa mode resonance frequency of 5.618 GHz and a power level of 10 dBm is generated by a signal generator (Windfreak SynthUSB3, SG). This signal is sent through a splitter (Mini-Circuits ZX10R-14-S+). The first output of the splitter is connected to a circulator (DiTom D3C2080), while the second output is connected to the local oscillator (LO) port of an IQ mixer (Marki Microwave MMIQ-0520L). To perform the switching of the IDT responses, we utilize a 10 kHz square wave with varying peak-to-peak voltages ($V_{\text{pp}}$) generated by a waveform generator (Keysight 33511B, WG). This poling signal is combined with the RF signal using a bias tee (Quantum microwave QMC-CRYODPLX-S0218NM), the output of which is sent to the device under test (DUT). The reflected microwave signal is routed to the RF port of the IQ mixer, and time-domain variations in the reflection signal are monitored by an oscilloscope (Tektronix TDS 2014B) connected to the in-phase port of the IQ mixer. 

To verify that the measured switching dynamics is not limited by the bandwidth of the setup, the output of the waveform generator after going through the bias tee is directly measured by the oscilloscope, which gives a time constant of 4 ns. On the other hand, the bandwidth of the intermediate frequency (IF) port is from 0 to 6 GHz. Additionally, the intrinsic linewidth of the measured Sezawa resonance is $\sim$60 MHz, corresponding to a cavity lifetime of 17 ns. It can be seen the response speed of the measurement is primarily limited by the lifetime of the SAW resonance which is a few times shorter than the time constants of the measured switching dynamics. Therefore, the switching time is dominated by the time required for the polarization reversing, and the obtained time-domain response reflects the dynamics of the domain switching.

\subsection{Modeling of the dynamics}
Ferroelectric domain switching has been widely studied in the past few decades, and it is commonly adopted that the switching involves domain nucleation and growth \cite{geler2022ferroelectric, shin2007nucleation}.
Different models have been proposed to describe the switching dynamics based on different pictures of the nucleation and domain growth process. A conventional model is the so-called Kolmogorov-Avrami-Ishibashi (KAI) model, which describes the change of polarization as \cite{jo2007domain}: 
\begin{equation}
    \Delta P(t)=A(1-e^{-(t/t_0)^n})
\end{equation}
where $t_0$ is a characteristic switching time, and $n$ is the effective dimensionality of the domain wall motion with $n=2$ for thin films. It assumes single nucleation and the growth of it to switch the whole film \cite{geler2022ferroelectric, jo2007domain}. However, it is found that it can only describe the initial fast switching but not the slow tail of the switching curve for some ferroelectric thin films \cite{gruverman2005direct}. The picture is not suitable for multi-domain BTO film since each domain can have its own nucleation and different nucleation time. Another model was proposed to take into account the statistic distribution of nucleation times, assuming the switching is limited by the nucleation time and the time for domain wall motion is negligible (since each domain is small). This is often referred to as the nucleation limited switching (NLS) model \cite{tagantsev2002non}. In the model, the switching is the response of an ensemble of independent areas with nucleation time $\text{log}~t_0$ follows a Lorentzian distribution with mean value of $\text{log}~t_{\text{mean}}$ and width of $w$ \cite{tagantsev2002non, jo2007domain}. The dynamics of the polarization can be described by \cite{geler2022ferroelectric, jo2007domain}:
\begin{equation}
    \Delta P(t)=A\left(\frac{1}{2}+\frac{1}{\pi}\text{arctan}\frac{\text{log}~t-\text{log}~t_{\text{mean}}}{w}\right)
\end{equation}
It is found that the trend of the switching dynamics obtained in the experiment can be qualitatively described by the above equation but with slight discrepancy. We attribute it to the fact that in the experiment we measure the changing of the depth of SAW resonance which has an indirect and nonlinear relation with the polarization of film. Further exploration of the experiment and modeling will be needed to better understand the domain switching in BTO thin films. 

To extract the time constant of the measured state switching process, we found the result can be empirically fitted with a two-time-constant exponential fitting:
\begin{equation}
    S(t)= A_0+A_1(1-e^{-t/\tau_1})+A_2(1-e^{-t/\tau_2})
\end{equation}
where $S$ represents the homodyne signal, and $\tau_1$ and $\tau_2$ are the time constants for the fast and slow processes, respectively. A good fitting can be obtained for various switching voltages as shown in Fig. \ref{fig:switching}f in the main text. The slower switching tail with $\tau_2$ can be attributed to a slower domain wall movement, which is related with grain boundaries, dislocations, oxygen vacancies, and surface and bound charges, which will require detailed study and modeling in the future \cite{li2019real}.

\begin{figure*}[h]
    \centering
    \includegraphics[width=15.5 cm]{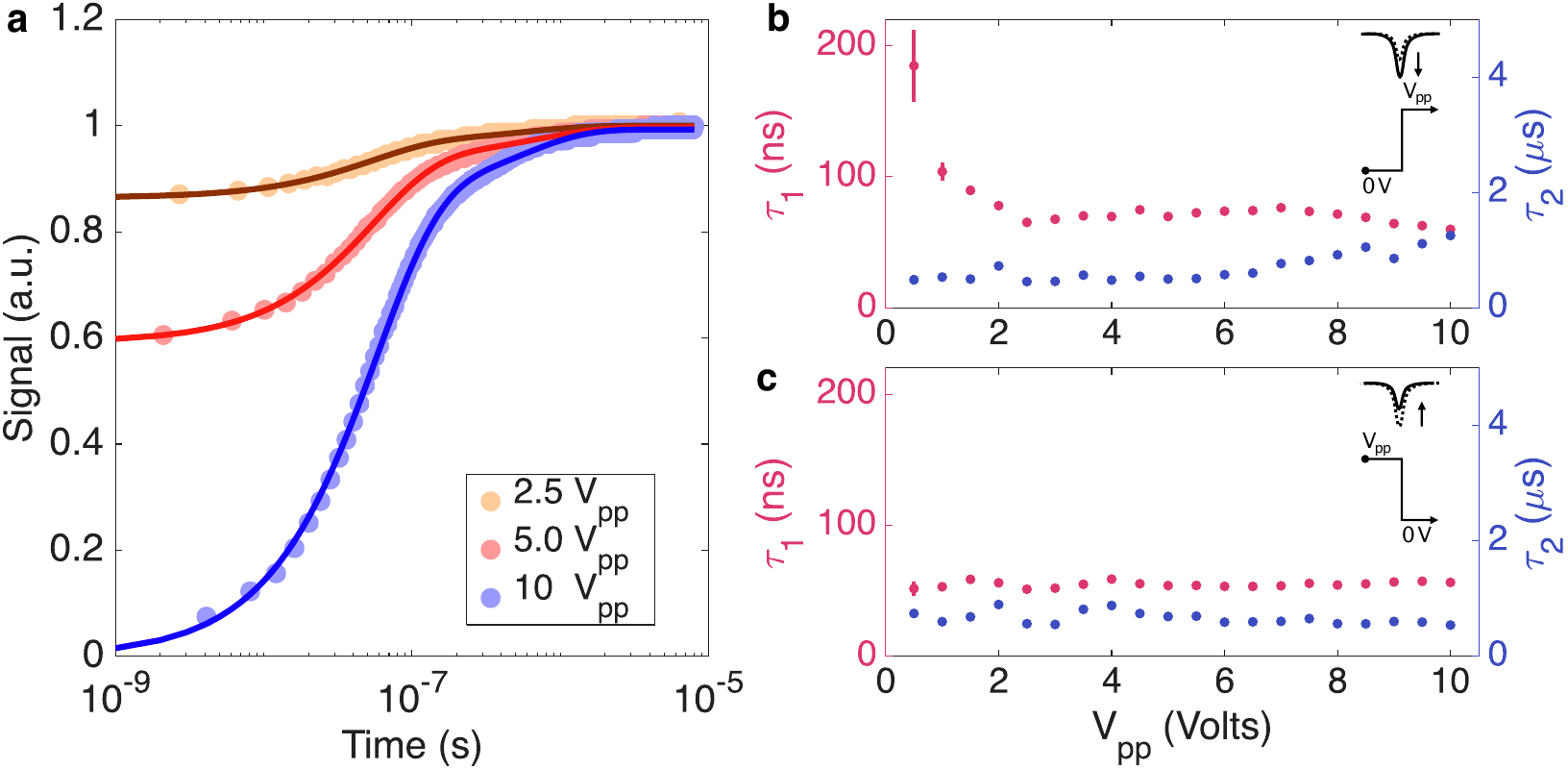}
    \caption{\textbf{Depolarization dynamics and voltage dependence } \textbf{a,} The time-domain dynamics of the depolarization process when switching from various $V_{\text{pp}}$ back to 0 V (see inset in \textbf{c}). Solid lines indicate empirical fittings to the experimental data using a two-time-constant exponential model. \textbf{b} and \textbf{c,} The extracted two time constants $\tau_1$ and $\tau_2$ for the poling and depolarization process, respectively. The error bars are 95\% CI of the fitting. The large CI at low voltages is due to the low voltage resolution of the oscilloscope. The insets in \textbf{b} and \textbf{c} illustrate the changing of the SAW resonance (top), and the applied switching signal (bottom) which correspond to the rising and falling edges of the square wave. }
    \label{Bandwidth} 
\end{figure*}

\subsection{Depolarization dynamics}
In the main text, the dynamics when switching from 0 V to a positive bias is presented, which corresponds to the poling of domains under an external electric field. As we apply a square wave in the experiment, the transient when the bias switches back to 0 V is also obtained. The changing of the measured signal is related with the depolarization process initiated by the depolarization field \cite{zhao2019depolarization}. The results for various $V_{\text{pp}}$ are presented in Fig. \ref{Bandwidth}a. A similar trend as the poling process is observed, which is also fitted with the two-time-constant exponential.

\subsection{Dependence of switching time on voltage}
Previous studies have shown that the switching time increases as the electric field decreases \cite{geler2022ferroelectric, jo2007domain}, which obeys the well-established Merz’s law \cite{merz1954domain}, an empirical relation stating that the switching time is exponentially dependent on the ratio of an activation field to the applied electric field. Here, the activation field, a material-dependent parameter, represents the energy barrier for domain switching. The fitted $\tau_1$ and $\tau_2$ for the poling process under a wide range of applied voltages are plotted in Fig. \ref{Bandwidth}b. It is observed that $\tau_1$ is nearly constant when the voltage is decreased to 2.5 V and starts increasing with lower voltages. Sub-100 ns of $\tau_1$ is kept until 1 V. On the other hand, the slow process $\tau_2$ is on the order of $\mu$s. The overall switching is dominated by the fast process as it has a relative magnitude ($A_1$) that is an order of magnitude larger than that of the slow process ($A_2$) from the fitting. The voltage dependence of the time constants for the depolarization process is shown in Fig. \ref{Bandwidth}c. Both of the $\tau_1$ and $\tau_2$ are kept constant during the course of the applied voltages, with average $\tau_1$ and $\tau_2$ of 55 ns and 0.7 $\mu$s, respectively. 

\section{Full 180\texorpdfstring{$^{\circ}$}{°} IDT rotation}
\begin{figure*}[t]
    \centering
    \includegraphics[width=10 cm]{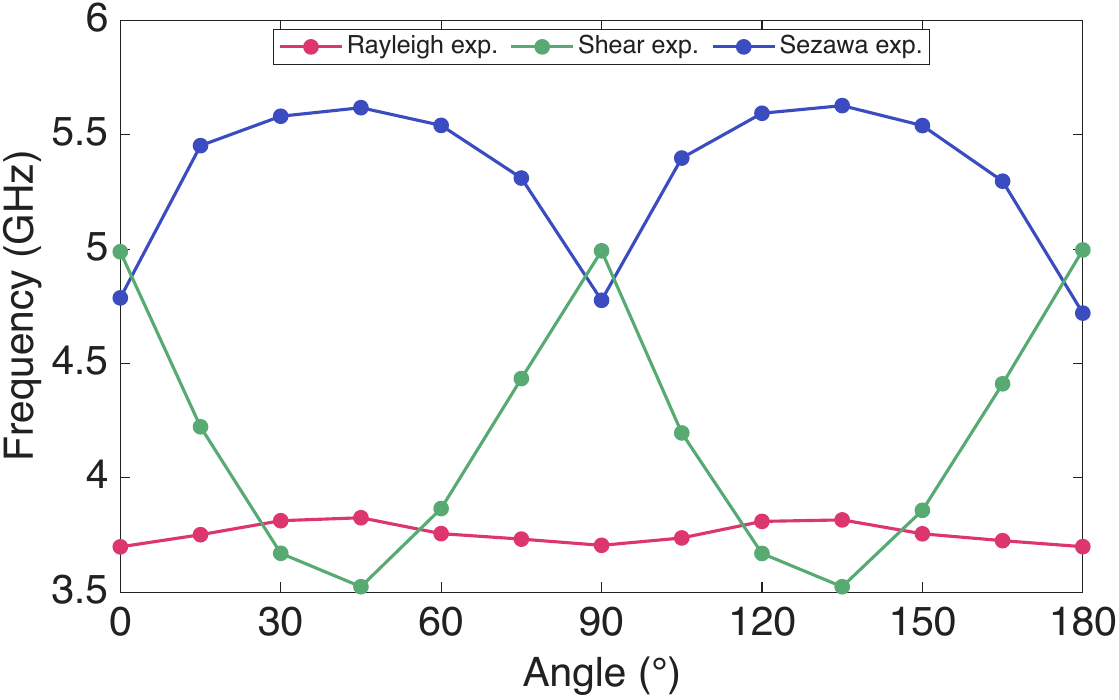}
    \caption{\textbf{Acoustic resonances for IDTs with $180^{\circ}$ rotation.} Angle dependence of the mechanical frequencies of Rayleigh (red), shear (green), and Sezawa (blue) modes from devices on Chip 1. $90^{\circ}$ periodicity can be observed, agreeing with the multi-domain BTO picture. }
    \label{5000Freq} 
\end{figure*}

A key property of the multi-domain model is the $90^{\circ}$ periodicity, due to the 4-fold symmetry of the domain distributions as illustrated in Fig. \ref{Schematics}. This is validated by measuring the frequency response of IDTs with a full $180^{\circ} $ rotation, as shown in Fig. \ref{5000Freq}. A $90^{\circ}$ periodicity of the acoustic resonances can be observed for all the three modes, which agrees with the multi-domain BTO picture. It also verifies the existence of $90^{\circ}$ and $270^{\circ}$ domains with equal populations with $0^{\circ}$ and $180^{\circ}$. Additionally, it is consistent with the single crystal assumption as polycrystalline would give a uniform response over orientation. 

\section{Multi-domain  mechanical and piezoelectric modeling}
The multi-domain BTO microstructure model is constructed as an ensemble of single domain elements, each possessing the same intrinsic elastic, dielectric, and piezoelectric material properties. The only difference among them is their in-plane polarization direction, which can be $0^\circ$, $90^\circ$, $180^\circ$, or $270^\circ$ relative to the laboratory frame.
The direction of the spontaneous polarization determines how the material matrices are rotated. Throughout the supplementary, we refer to these individual domains as the $0^\circ$/$90^\circ$/$180^\circ$/ $270^\circ$ domains. By assembling these individual domains randomly into a mosaic pattern, we capture the microscopic heterogeneity inherent to BTO films \cite{abel2019large}.
 
To determine the macroscopic effective response of the multi-domain film, we first express the compliance, dielectric, and piezoelectric properties in Voigt notation, which allows us to reduce these tensors into their matrix counterparts. We apply two homogenization approaches to extract the material coefficients. First, the Voigt-Reuss-Hill (VRH)\cite{hill1952elastic} averaging provides an empirical estimate for the compliance matrix. Secondly, a full finite-element method (FEM) simulation of the coupled piezoelectric equations is performed on randomized mosaic distributions, enabling numerical extraction of the piezoelectric matrices. Ensemble averaging over multiple iterations of these random mosaics obtains a statistically effective material constants that takes into account the random variations of the microstructure.

\subsection {Compliance and piezoelectric matrices rotations} 

Considering an original coordinate system  $x_{1}$, $x_{2}$, $x_{3}$ and a rotated coordinate system $x_{1}'$, $x_{2}'$, $x_{3}'$ . The orientation of the new axes relative to the original frame can be described by the directional cosines \cite{ting1996anisotropic,cady1946piezoelectricity}

\[
\begin{array}{c|ccc}
   & x_1' & x_2' & x_3' \\
\hline
x_1 & \alpha_1 & \beta_1 & \gamma_1 \\
x_2 & \alpha_2 & \beta_2 & \gamma_2 \\
x_3 & \alpha_3 & \beta_3 & \gamma_3 \\
\end{array} \quad
\begin{bmatrix}
\alpha_1 & \beta_1 & \gamma_1 \\
\alpha_2 & \beta_2 & \gamma_2 \\
\alpha_3 & \beta_3 & \gamma_3
\end{bmatrix}=
\begin{bmatrix}
\vec{x}_1 \cdot \vec{x}_1' & \vec{x}_1 \cdot \vec{x}_2' & \vec{x}_1 \cdot \vec{x}_3' \\
\vec{x}_2 \cdot \vec{x}_1' & \vec{x}_2 \cdot \vec{x}_2' & \vec{x}_2 \cdot \vec{x}_3' \\
\vec{x}_3 \cdot \vec{x}_1' & \vec{x}_3 \cdot \vec{x}_2' & \vec{x}_3 \cdot \vec{x}_3'
\end{bmatrix}
\]
The compliance matrix $\mathbf{s'}$ in the new coordinate system would then be expressed by the matrix multiplication of the compliance $\mathbf{s}$ in the original coordinate system and transform matrices $\mathbf{{A}_t^{-1}}$and $\mathbf{{A}^{-1}}$.
\begin{align}
\mathbf{s}' &= \mathbf{A}_t^{-1} \mathbf{s} \mathbf{A}^{-1} 
\end{align}
where  $\mathbf{{A}_t^{-1}}$ is the transpose of $\mathbf{{A}^{-1}}$
\begin{align}
\textbf{A}^{-1} =
\begin{bmatrix}
\alpha_1^2 & \beta_1^2 & \gamma_1^2 & 2\beta_1 \gamma_1 & 2\gamma_1 \alpha_1 & 2\alpha_1 \beta_1 \\
\alpha_2^2 & \beta_2^2 & \gamma_2^2 & 2\beta_2 \gamma_2 & 2\gamma_2 \alpha_2 & 2\alpha_2 \beta_2 \\
\alpha_3^2 & \beta_3^2 & \gamma_3^2 & 2\beta_3 \gamma_3 & 2\gamma_3 \alpha_3 & 2\alpha_3 \beta_3 \\
\alpha_2 \alpha_3 & \beta_2 \beta_3 & \gamma_2 \gamma_3 &
\beta_2 \gamma_3 + \beta_3 \gamma_2 &
\gamma_2 \alpha_3 + \gamma_3 \alpha_2 &
\alpha_2 \beta_3 + \alpha_3 \beta_2 \\
\alpha_3 \alpha_1 & \beta_3 \beta_1 & \gamma_3 \gamma_1 &
\beta_3 \gamma_1 + \beta_1 \gamma_3 &
\gamma_3 \alpha_1 + \gamma_1 \alpha_3 &
\alpha_3 \beta_1 + \alpha_1 \beta_3 \\
\alpha_1 \alpha_2 & \beta_1 \beta_2 & \gamma_1 \gamma_2 &
\beta_1 \gamma_2 + \beta_2 \gamma_1 &
\gamma_1 a_2 + \gamma_2 \alpha_1 &
\alpha_1 \beta_2 + \alpha_2 \beta_1
\end{bmatrix}
\end{align}
Similarly, the piezoelectric matrix $\mathbf{d'}$ on the new coordinate system would be
\begin{align}
\mathbf{d}' &= \mathbf{a} \mathbf{d} \mathbf{A}^{-1} 
\end{align}
where,
\begin{align}
\mathbf{a} =
\begin{bmatrix}
\alpha_1 & \alpha_2 & \alpha_3 \\
\beta_1 & \beta_2 & \beta_3 \\
\gamma_1 & \gamma_2 & \gamma_3
\end{bmatrix}  
\end{align}

\begin{figure*}[t]
    \centering
    \includegraphics[width=15cm]{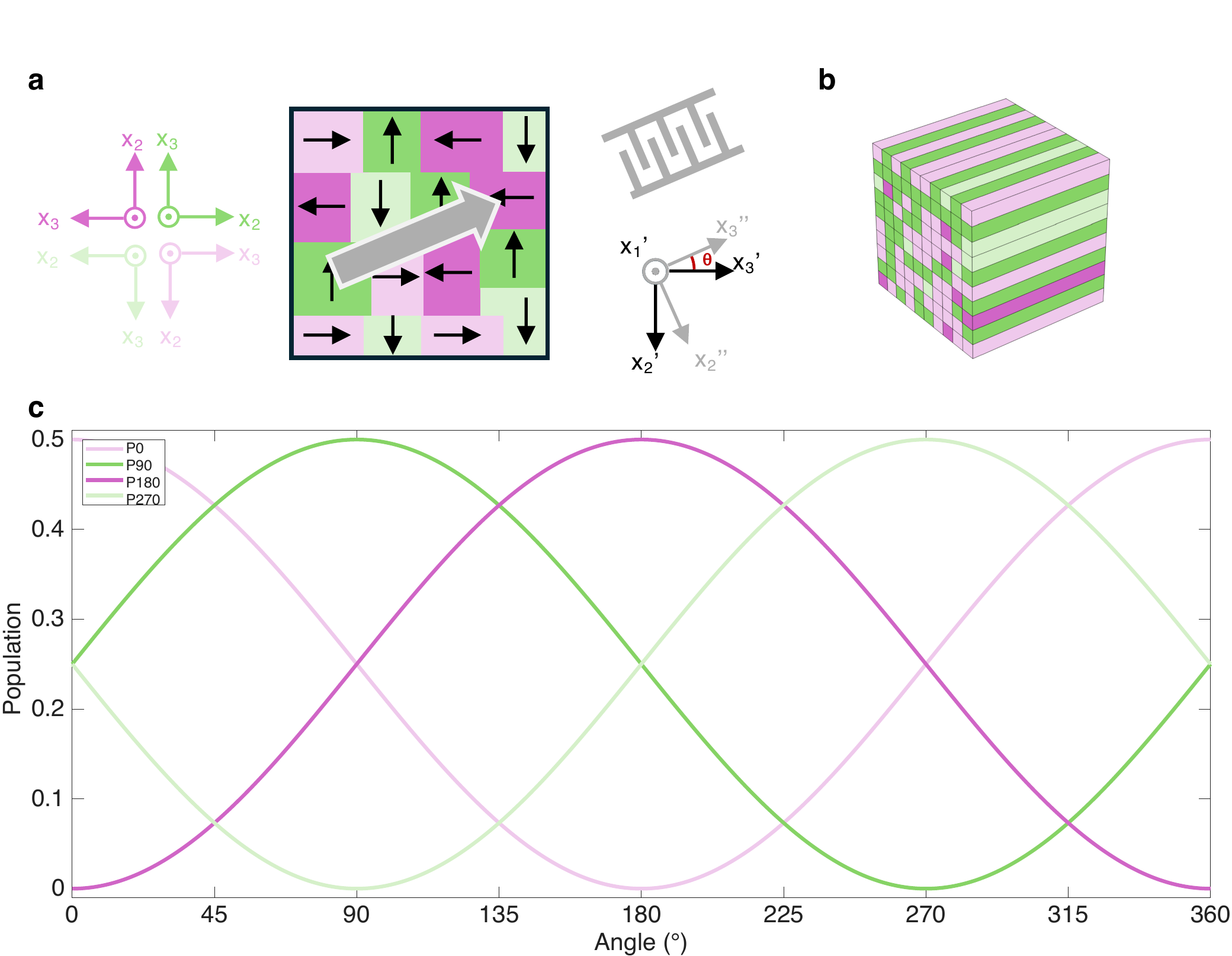}
    \caption{\textbf{Multi-domain BTO microstructure model.}  \textbf{a,} Schematic of coordinate definitions used in this work. Single domain coordinates $x_{1}$, $x_{2}$, $x_{3}$ are specific to each domain, dependent on the polarization directions. They are colored to match the domains in the schematic to the right. 
    Material homogenization over the four domains are done in the lab frame (black) $x_{1}'$, $x_{2}'$, $x_{3}'$. Finally, the coordinate is transformed from the lab frame to the the rotated device frame (gray) $x_1''$ $x_2''$ $x_3''$ in the numerical modeling of our BTO SAW resonators. \textbf{b,} FEM model used in evaluating effective piezoelectric values. Each colored element represents each of the domain. The position of each domain is generated randomly during each simulation. \textbf{c,} Domain population adopted in the simulation as a function of device orientation. }
    \label{Schematics} 
\end{figure*}

\subsection {Projecting material coefficients onto laboratory and device frame} 
For tetragonal BTO, the crystallographic 1 and 2 axes are symmetry-degenerate, while the 3-axis (c-axis) coincides with the spontaneous polarization direction. In our multi-domain construction, we fix the crystallographic 1-axis to be out-of-plane for all domains. The remaining 2 and 3 axes are then allowed to rotate within the film plane, according to the four possible in-plane polarization variants ($0^\circ$, $90^\circ$, $180^\circ$, $270^\circ$). The original coordinate axis for each domain are shown of Fig. \ref{Schematics}a, with the same color coding as the domain. 
%% maybe in the figure color label what is 0, 90, 180, 270 domain
We define the laboratory frame (black) having the 1-axis out of plane and its 3-axis parallel to the polarization of $0^\circ$ domains. All elastic and piezoelectric material homogenization calculations (VRH and FEM based) are performed in this lab frame. Prior to calculating the effective electromechanical $k^2_{\text{eff}}$ and mechanical resonance frequency, the homogenized material matrices undergo a second coordinate transform from the laboratory frame into the device frame (gray) defined by the orientation of the IDT.

\subsection {Material homogenization methods} 

\paragraph{Domain Population}  

Depending on the IDT orientation, the effective electric field projected along each domain's c-axis is varied , which leads to an angle-dependent domain population. 
We assume that $0^\circ$/$90^\circ$/$180^\circ$/ $270^\circ$ domains have populations of $P_{0^{\circ}}/P_{90^{\circ}}/P_{180^{\circ}}/P_{270^{\circ}}$ respectively, and $P_{0^{\circ}}+P_{90^{\circ}}+P_{180^{\circ}}+P_{270^{\circ}}=1$. Prior to any poling, we assume the population of the four domains is equally distributed as $P_{0^{\circ}}=P_{180^{\circ}}=P_{90^{\circ}}=P_{270^{\circ}}=0.25$. Since electrical poling can only flip the polarization by $180^{\circ}$, we further assume $P_{0^{\circ}}+P_{180^{\circ}}=P_{90^{\circ}}+P_{270^{\circ}}=0.5$. At $0^{\circ}$, it is assumed that the device is fully poled along $0^{\circ}$ direction with $P_{0^{\circ}}=0.5$ and $P_{180^{\circ}}=0$. Following Ref. \cite{abel2019large}, $P_{0^{\circ}}/P_{90^{\circ}}/P_{180^{\circ}}/P_{270^{\circ}}$ can be estimated by a sinusoidal function of device orientation, as plotted in Fig. \ref{Schematics}c.

\paragraph{VRH homogenization for elastics.}  After performing coordinate transforms of the stiffness and compliance tensors for each BTO single domain into the laboratory frame, we homogenized the elastic response of the multi-domain ensemble using the Voigt-Reuss-Hill (VRH) method. The Voigt and Reuss represent two extreme cases with iso-strain and iso-stress assumption, respectively. They give the upper and lower bounds of stiffness for a heterogeneous medium: the Voigt bound corresponds to the volume average of the stiffness matrices (under the iso-strain assumption), whereas the Reuss bound corresponds to the volume average of the compliance matrices (iso-stress). The VRH approximation of the effective stiffness is then taken as the arithmetic mean of these two extreme cases, and provides an empirically reasonable estimate of the elastic properties of a multi-domain film.

\paragraph{FEM homogenization of piezoelectric coefficients.} 

Homogenization of the effective piezoelectric coefficients is performed using FEM simulations. We model the response of our films with a unit cell constituted by a 10$\times$10 mosaic ensemble of single domain components Fig. \ref{Schematics}b. This is chosen to reflect the relative physical length scales as in-plane BTO domains are approximately 100 nm, while the IDT finger pitch is on the order of 1 $\mu$m. Therefore, this unit cell represents the piezoelectric interaction over one wavelength.  
To probe the effective piezoelectric coefficient $d_{ij,\text{eff}}$, we apply 6 stress loads independently to the unit cell, with each only activating one Voigt stress component ($T_{1}$=1 Pa, $T_{2}$=1 Pa, $T_{3}$=1 Pa, $T_{4}$=1 Pa, $T_{5}$=1 Pa, and $T_{6}$=1 Pa), while keeping all other stresses to be 0. Short circuit electrical boundary conditions (zero electrical potential difference across opposite faces of the unit cell) are enforced for all load cases. %The resulting volume-averaged electric displacement fields $\langle D_i\rangle$ are then used to obtain each effective piezoelectric coefficient 
The resulting effective piezoelectric coefficient is then calculated as $d_{ij,\text{eff}}= \frac{\langle D_i\rangle}{\langle T_j\rangle}$ via the strain charge relation. 
\begin{align}
\begin{bmatrix}
D_1 \\ D_2 \\ D_3
\end{bmatrix}
=
\begin{bmatrix}
d_{11} & d_{12} & d_{13} & d_{14} & d_{15} & d_{16} \\
d_{21} & d_{22} & d_{23} & d_{24} & d_{25} & d_{26} \\
d_{31} & d_{32} & d_{33} & d_{34} & d_{35} & d_{36} \\
\end{bmatrix}
\begin{bmatrix}
T_1 \\ T_2 \\ T_3 \\ T_4 \\ T_5 \\ T_6
\end{bmatrix}
+
\begin{bmatrix}
\varepsilon_{11} & 0 & 0 \\
0 & \varepsilon_{22} & 0 \\
0 & 0 & \varepsilon_{33}
\end{bmatrix}
\begin{bmatrix}
E_1 \\ E_2 \\ E_3
\end{bmatrix}
\end{align}

The choice of the above unit cell with 10$\times$10 random domain distribution is for the sake of simulation time. In reality, the size of an IDT is much larger, which could contain orders of magnitude more domains. The net result is a converged and stabilized piezoelectric response with little variation from device to device. To account for this and to remove the noise due to randomization in the simulation, we take the average over N = 10 independent Monte-Carlo realization of the microstructure, each with randomized domain arrangements but same domain population. The choice of N = 10 is to account for the tradeoff between convergence of the homogenization and computation efficiency. Compared to a much larger iteration number (N = 150), the numerical spread of $d_{33,\text{eff}}$ at N = 10 is roughly within 2.5\% of the converged value, which indicates that N = 10 provides a stable solution and saves computation resources for our homogenization scheme. This variation due to randomization will be considered when calculating fitting confidence interval later.

This method aligns with the notion of the representative volume element (RVE) \cite{kanit2003determination} -- defined as the smallest ensemble of elements that reproduces the macroscopic behavior of a heterogeneous medium. The RVE is not defined purely by the number of elements, but rather by the requirement that effective properties converges with acceptable variance. Under this framework, smaller unit cells can also be representative of the the RVE by averaging over a sufficient number of independent instances. Therefore, our use of N = 10 Monte Carlo iteration gives homogenized material coefficients that are representative of device-scale behavior.

%\begin{figure*}[ht]
   % \centering
   % \includegraphics[width=13cm]{Figures/d33converge.pdf}
    %\caption{\textbf{Convergence of RVE for different Monte Carlo iterations.} The y-axis is normalized  }
   % \label{MC Convergence} 
%\end{figure*}

\begin{figure*}[t]
    \centering
    \includegraphics[width=\textwidth]{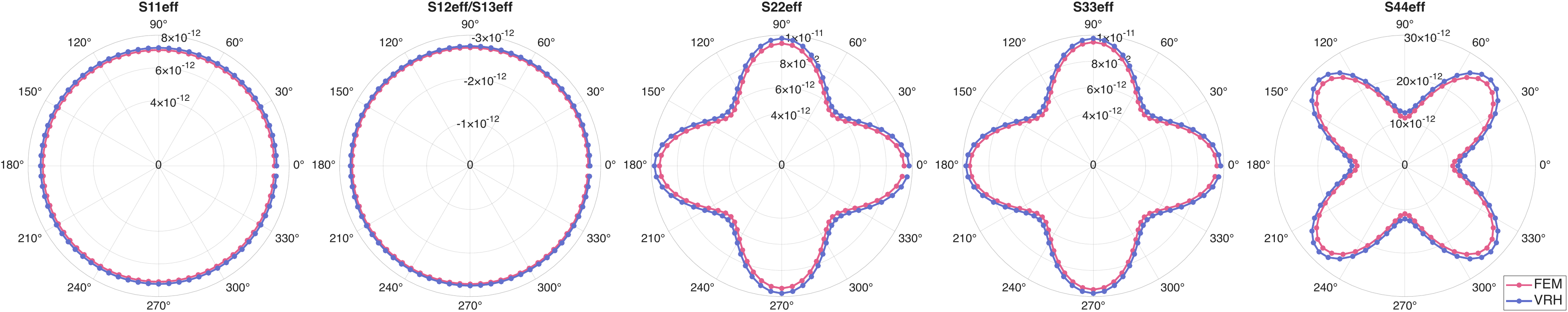}
    \caption{\textbf{Homogenized elastic constant comparison between FEM simulation and VRH calculations}  }
    \label{VRH MC PolarPlot} 
\end{figure*}

Note that the FEM mosaic model described above can also be used to calculate the effective elastic coefficients by following similar procedure, as shown in Fig. \ref{VRH MC PolarPlot}. It is interesting to see that there is a good agreement between analytical VRH method and FEM simulation, which cross validates the two method. However, we opted for the analytical solution for better computation efficiency.

\paragraph{$k_{\text{eff}}^2$ simulation of IDT devices} 
Taking the homogenized elastic and piezoelectric coefficients from the VRH and FEM analysis, we perform a frequency domain analysis in an FEM model of the IDT, which considers an IDT finger pair with periodic boundary conditions. $k_{\text{eff}}^2$ is extracted via BVD model fitting of the simulated $Y_{11}$ response.

\subsection {Material coefficient and confidence interval extraction} 

\begin{figure*}[t]
    \centering
    \includegraphics[width=15cm]{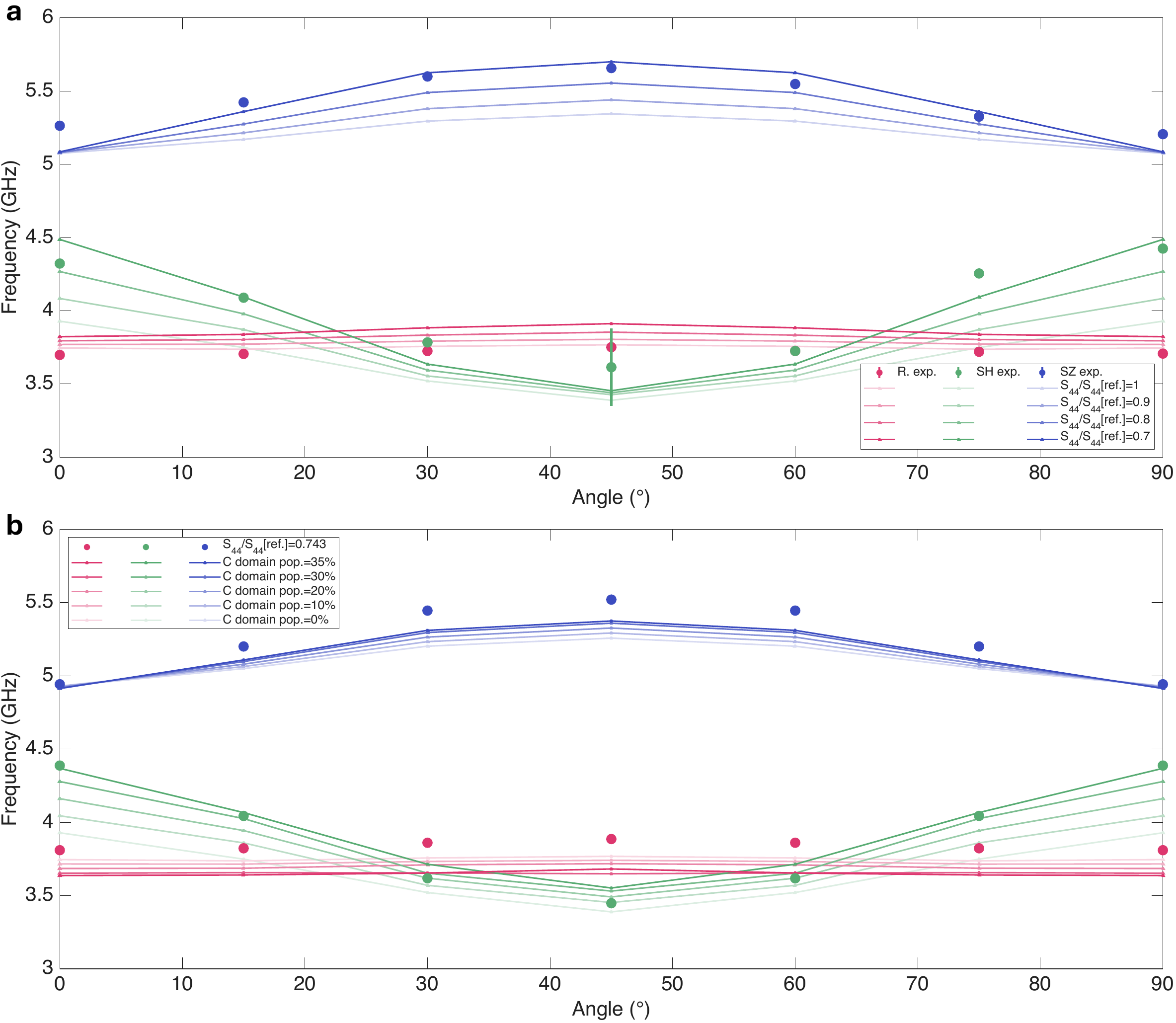}
    \caption{\textbf{Dependence of resonant frequency on compliance $s_{44}$ and c-domain population.} \textbf{a,} Comparison between experiment (dots) and simulations with various $s_{44}$ for the Rayleigh (red), shear (green), and Sezawa (blue). The $s_{44}$ is normalized to the value in Ref. \cite{zgonik1994dielectric,meng2010dft}. Decreasing of $s_{44}$ (from light to dark) shows monotonic increasing of frequency for the three modes, and matches the experimental trend for Sezawa and shear modes. \textbf{b,} Simulated resonant frequency under various c-domain population, while the $s_{44}$ is taken from Ref. \cite{zgonik1994dielectric}. It is compared with the results obtained by using the fitted $s_{44}$ value (dots). With the increase in c-domain population, the shear mode's frequency increases and approaches the results from $s_{44}$ fitting.}
    \label{S44 Tuning & C domain} 
\end{figure*}

The extraction of material properties is performed using a 2-stage analysis. In the first stage, the elastic property of the BTO is fitted to match the simulated mechanical resonant frequency with the experiments. As will be discussed in the next section, the shear compliance coefficient $s_{44}$ serves as the only fitting parameter while others are taken from Ref. \cite{zgonik1994dielectric}. For each IDT orientation, we compute the effective elastic matrix using the VRH method. The $s_{44}$ is fitted by matching the angular trend of the simulated and experimental resonance frequency of the shear mode, as shown in Fig. \ref{S44 Tuning & C domain}a. We extract $s_{44}=12.18\times 10^{-12}\pm0.33\times 10^{-12} $[1/Pa] with 95\% confidence interval. This fitted elastic constant, alongside other compliance terms found in Ref. \cite{zgonik1994dielectric, meng2010dft} are then taken as fixed inputs of the subsequent analysis.  

In stage 2, the compliance matrix (from stage 1), the dielectric matrix (taken from capacitance measurement), and the piezoelectric coefficients ($d_{31}$, $d_{31}$ and $d_{15}$), which serve as fitting parameters, are supplied to the 2D mosaic FEM simulation. The obtained effective piezoelectric matrix along with the VRH effective elastic matrix is fed into the IDT FEM model to simulate the Sezawa mode $k_{\text{eff}}^2$. By fitting the angular dependency of the Sezawa mode electromechanical coupling coefficient, we extract the piezoelectric coefficients with 95\% confidence interval (CI) as: $d_{31}=-47.54\pm0.85$ pC/N, $d_{33}=78.89\pm1.99$ pC/N and $d_{15}=204.91\pm4.10$ pC/N.

To capture the full uncertainty of our model, two additional noise sources are also analyzed. First, the error due to the variation across different independent instances of the Monte Carlo microstructure should be considered. To quantify this effect, we generate numerous independent mosaics, and examine the statistical spread of the resulting $k_{\text{eff}}^2$. Across all angles, this translates to a maximum variation of the extracted piezo-coefficient about $3\%$. We include this as an extra standard error term induced by the Monte Carlo variation in the final reported CI. 
Additionally, the uncertainty from the $s_{44}$ estimation is propagated into $d_{31}$, $d_{31}$ and $d_{15}$ using the Murphy-Topel 2 stage estimator \cite{murphy2002estimation,hardin2002robust}. Finally, we report the total 95\% confidence interval  for the piezo-coefficients as $d_{31}=-47.54\pm 4.83$ pC/N, $d_{33}=78.89\pm7.82$  pC/N and $d_{15}=204.91\pm 31.55$ pC/N.

\subsection {Justification for $s_{44}$ tuning} 
As shown in the main text, unlike the Rayleigh and Sezawa modes, the frequency of the shear mode decreases as the IDT orientation approaches 45$^{\circ}$, which is primarily caused by the angular dependence of $s_{44,\text{eff}}$. It can be seen from the angular trend of each compliance component shown in Fig. \ref{VRH MC PolarPlot} that only the $s_{44,\text{eff}}$ maximizes at 45$^{\circ}$. Since the acoustic velocity is proportional to $1/\sqrt{s}$, this leads to the decreasing of the shear mode's resonance at 45$^{\circ}$. Further analytical analysis of the homogenized $s_{44}$ shows that it is dominated by the single domain $s_{44}$. 

To confirm that fitting only the $s_{44}$ term is sufficient in fitting both Rayleigh, Sezawa, and shear modes, we compare the FEM simulated frequency response as a function of $s_{44}$ while keeping all other compliance coefficient constant, as shown in Fig. \ref{S44 Tuning & C domain}. By adopting the original compliance reported in Ref. \cite{zgonik1994dielectric}, both Rayleigh and Sezawa modes show good agreement with the experiment, while the shear mode is far below the experiment (see the curves with the lightest color). This suggests it is primarily $s_{44}$ that deviates from Ref. \cite{zgonik1994dielectric}.
As $s_{44}$ decreases, we see an increase in frequency for Sezawa mode at $45^\circ$ and shear mode at $0^\circ$ respectively, both agreeing with the trend seen in the experiment. Therefore, we choose to fit $s_{44}$ in our numerical modeling. A good match with the experiment can be obtained by scaling the $s_{44}$ by 0.743 relative to the value given in Ref. \cite{zgonik1994dielectric}.

While further study is needed to explain the deviation of $s_{44}$, some preliminary analysis hints the relation with the microstructure of BTO thin films. It is noticed from the previous study of RF-sputtered BTO films that there exists a decent portion of c-domains in a-domain BTO films and also polycrystalline regions near the surface \cite{posadas2021thick}. These defects are not captured in our modeling, which assumes pure a-domain and single crystal. The impact of the c-domains on the effective elasticity can be analyzed similarly via the VRH method. Here, we adopt the original compliance from Ref. \cite{zgonik1994dielectric} while changing the population of c-domain. Figure \ref{S44 Tuning & C domain}b shows the comparison of the resonant frequency for various c-domain population with the results obtained by fitting the $s_{44}$. As the c-domain population increases, the shear mode's frequency increases and approaches the $s_{44}$ fitting. A good matching is achieved with c-domain taking up 35\% of the total population. 
This is consistent with domain analysis on 300 nm RF-sputtered BTO a-axis films \cite{posadas2021thick}, where only 51\% of the population is purely a-domain. We emphasize that the real microstructure of mixed a/c-domain films is complex, including domain tilt/twinning and local texture. These features cannot be fully captured by our simplified 2D Monte-Carlo mosaic model, which leaves room for more detailed future studies that incorporate realistic morphology and electro-mechanical coupling. Nevertheless, our proposed FEM microstructure simulation platform can serve as a platform for modeling these more sophisticated microstructure. 

\subsection{Orientation dependence of $d_{33,\text{eff}}$}
\begin{figure*}[ht]
    \centering
    \includegraphics[width=10cm]{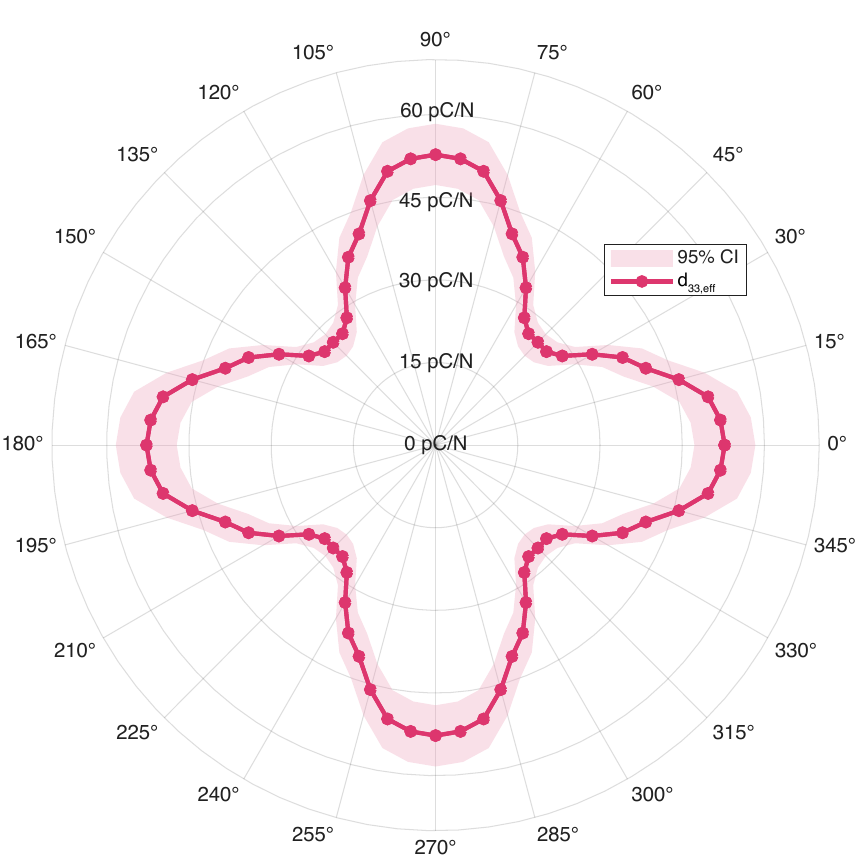}
    \caption{\textbf{Orientation dependence of $d_{33,\text{eff}}$.} The shaded region is the 95\% CI for the fitted $d_{33,\text{eff}}$. }
    \label{d33eff} 
\end{figure*}

In the main text, we gave the fitted $d_{33,\text{eff}}$ at 0$^{\circ}$. Here, the orientation dependence of the $d_{33}^{\text{eff}}$, alongside its $95\%$ confidence interval is shown in Fig. \ref{d33eff}. Similar to the effective compliance, the piezoelectric coefficient exhibits a $90^\circ$ periodicity, and minimizes at 45$^{\circ}$. 

\section{Cryogenic measurement and calibration}
\subsection{Measurement setup}
\begin{figure*}[t]
    \centering
    \includegraphics[width=5cm]{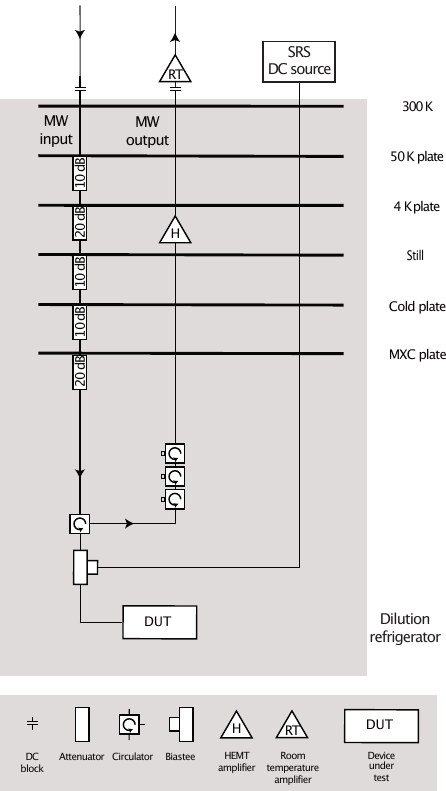}
    \caption{\textbf{Cryogenic measurement setup.} The schematic showing the wiring inside the dilution fridge.}
    \label{CryoSetup} 
\end{figure*}

Cryogenic measurements are performed with the setup shown in Fig. \ref{CryoSetup}. Our device under test (DUT) is wire-bonded to a printed circuit board (PCB) and enclosed within a copper box, which is mounted on the mixing chamber stage of a He-3/He-4 dilution refrigerator. The input microwave is routed through the input line and attenuated by 70 dB. The reflected signal from the DUT is routed through a circulator bank in the output line and subsequently amplified using a high electron mobility transistor amplifier (LNF-LNC0.3-14B HEMT) at the 4-k stage, followed by a room temperature amplifier. The input and output microwaves are connected to port 1 and 2 of a vector network analyzer (VNA) to measure the $S_{11}$ spectrum of the IDT. 

\subsection{Calibration method}
The cryogenic characterization is necessarily more limited in scope than the room-temperature study. At room temperature, a probe station allows many devices co-fabricated on the same chip to be measured rapidly, whereas each cryogenic measurement requires the device to be wire-bonded and packaged, limiting the number that can be characterized per cooldown. In addition, extracting the admittance and $k^2_{\text{eff}}$ directly, as done at room temperature, would require in-situ de-embedding and short-open-load (SOL) calibration of the cryogenic wiring, which the present setup is not equipped to perform; this motivates the self-calibrated $\gamma_e$ scaling used here.

The linewidth of the measured device (tens of MHz) is comparable to the free spectral range (FSR) of the parasitic standing wave modes from the dilution refrigerator wiring. These FSR ripples overlap with our device resonances in the $S_{11}$ spectrum, which necessitates careful calibration. Due to presence of circulators in our setup, a standard SOL microwave calibration is only possible via a specific VNA setup that needs to rewire the internal amplifiers of the VNA, as discussed in Ref. \cite{wang2021cryogenic}. To overcome this limitation, we implement a zero-delay-through calibration, a method in which the measured  $S_{11}$ is normalized to a reference $S_{11}$ trace along the same microwave path but without the DUT. However, it is not realistic to connect and disconnect the device frequently inside the fridge which would introduce excessive heating. Therefore, we utilize the ferroelectric property of the BTO that the piezoelectric response can be switched off near 0 V biasing, which only contains the background responses including cabling resonances, losses and gains of the input and output lines. We then normalize the $S_{11}$ spectrum to this background to obtain the calibrated piezoelectric response. It should be noted that the coercive field is temperature dependent which means the bias at which the piezo-response is zero varies with the temperature. Thus, we swept the biasing to find the coercive field at each temperature.

\subsection{Piezoelectric constant scaling at cryogenic temperature}
Since we use the zero-delay-through method which is self-calibrated, we cannot obtain the admittance of the IDT and thus fit for $k^2_{\text{eff}}$. Therefore, we estimate the piezoelectric response through the scaling of the external coupling coefficient $\gamma_{e}$ in the main text, which is proportional to the effective electromechanical coupling constant defined as \cite{blesin2021quantum, anderson2026high}:
\begin{equation}
    k_{\text{eff}}^2\propto K^2=\frac{e_{\mathrm{eff}}^2}{c_{\mathrm{eff}}\varepsilon_{\text{eff}}}
\end{equation}
where $K^2$ represents the material intrinsic electromechanical coupling factor, $c_{\text{eff}}$ and $\varepsilon_{\text{eff}}$ are the effective stiffness and permittivity, respectively. $e_{\mathrm{eff}}$ is the effective piezoelectric stress constant, which is related with $d_{\mathrm{eff}}$ through the constitutive relation $e_{\mathrm{eff}}=d_{\mathrm{eff}} c_{\text{eff}}$ \cite{tian2024}. Therefore, the external coupling scales as a function of effective piezoelectric constant, stiffness and permittivity as:
\begin{equation}
    \gamma_{e}\propto\frac{d_{\mathrm{eff}}^2c_{\mathrm{eff}}}{\varepsilon_{\text{eff}}}
\end{equation}
Comparing cryogenic measurements with those measured at room temperature, we observe a negligible increasing (\textless 3\%) in Sezawa mode resonant frequency, indicating that the effective stiffness and compliance are kept nearly constant during the cool down measurements. Due to the self-calibration method, we cannot extract the intrinsic capacitance $C_0$ of the IDT and thus the relative permittivity as a function of temperature. Previous work has studied the permittivity of BTO thin film down to 4 K \cite{eltes2020integrated}, which reported a roughly 1.7 times reduction in cryogenic dielectric constant compared to room temperature. Taking this into account the 4.6 times reduction in $\gamma_{e}$, we estimate a 2.8 times reduction of the effective $d_{33,\text{eff}}$ ($\sim$19 pC/N). 

It should be noted that in Ref. \cite{eltes2020integrated}, the reduction in the dielectric constant was observed to decrease as the electric field increased. Therefore, in our experiment with an electric field of 33 MV/m, we would expect a smaller reduction in the dielectric constant compared to that measured at 20 MV/m in Ref. \cite{eltes2020integrated}. In the extreme scenario with no dielectric reduction, the effective piezoelectric constant would be approximately 25 pC/N. Thus, a 2-3 times reduction of piezoelectric coefficient in cryogenic temperature is observed which agrees with the reduction of electro-optic coefficient measured in Ref. \cite{eltes2020integrated}. In the future, the piezoelectric coefficient at cold can be directly measured through better microwave calibration such as using cryogenic probe station with standard SOL calibration substrate \cite{russell2012cryogenic}, cryogenic single-port calibration \cite{wang2021cryogenic}, or on-board calibration \cite{shirachi2025board}.

\section{Fundamental Rayleigh mode}
%Due to the relatively low tetragonal phase Curie temperature of BTO  ($T_{c}\approx 120^\circ C$), continuous electric poling is necessary to maintain stable polarization at room temperature. This poling is achieved using a periodic poling setup, where a direct current (DC) bias is applied to the IDT bus.  This results in the characteristic acoustic wavelength $\lambda=\Lambda/2$ for the Rayleigh and Sezawa modes. 

\begin{figure*}[ht]
    \centering
    \includegraphics[width=\textwidth]{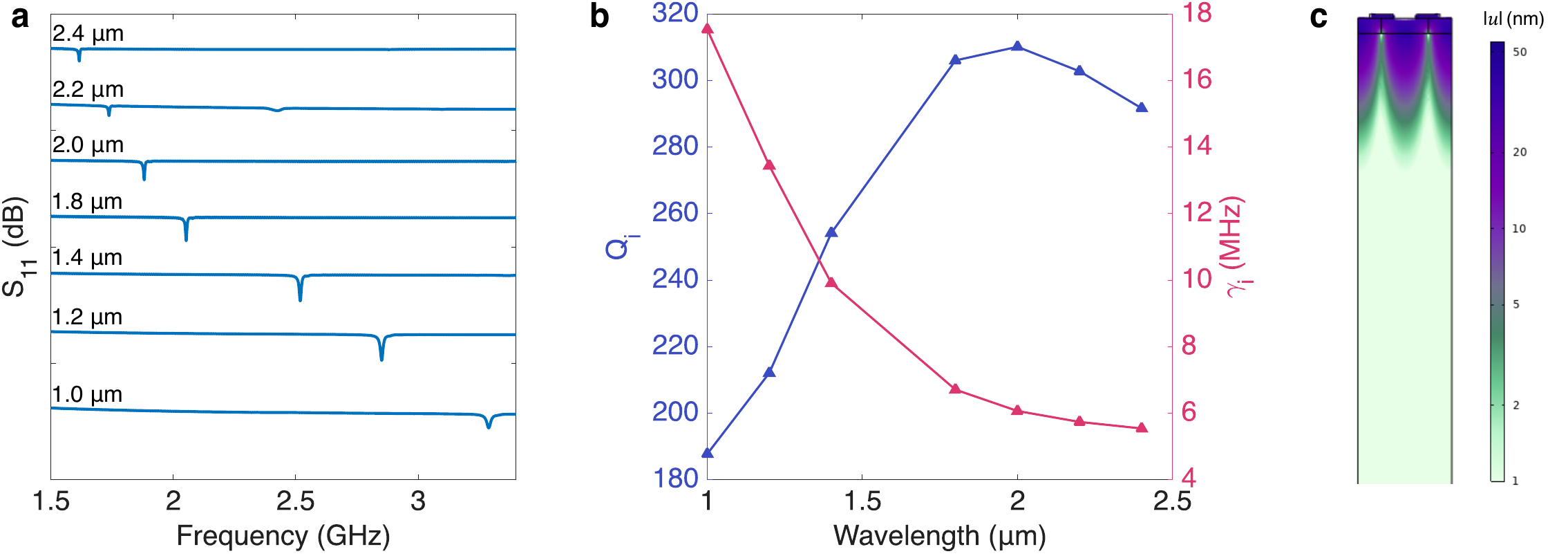}
    \caption{\textbf{Intrinsic Q of the fundamental Rayleigh mode.} \textbf{a,} $S_{11}$ spectra for IDTs with various IDT pitches as labeled. The spectra is shifted vertically for better illustration and the spacing between y-ticks is 10 dB. The measurements are taken at biasing of 0 V and highlight the acoustic response of the fundamental Rayleigh mode.  \textbf{b,} Intrinsic $Q_{i}$ (blue) and linewidth $\gamma_{i}$ (red) of the fundamental Rayleigh mode as a function of acoustic wavelength. Since it's fundamental mode, the wavelength equals to the IDT period. \textbf{c,}  Numerical simulation of the fundamental Rayleigh mode profile. The color bar is the magnitude of the displacement in logarithm scale. Compared to the Sezawa mode profile in Fig. \ref{Fig:bulk mode}b, the absence of BAW modes explains why the fundamental modes exhibit higher intrinsic quality factors. }
    \label{Fundamental} 
\end{figure*}

In addition to the modes studied in the main text that are compatible with the periodically poled IDT, we also observe acoustic resonances at lower frequencies that occur when the applied bias is near zero. These modes are identified as fundamental acoustic modes, with wavelengths equal to the IDT period. The existence of these modes originates from the weak remnant polarization, leading to a significantly weaker external electromechanical coupling. Figure \ref{Fundamental}a shows the spectra of the fundamental Rayleigh mode measured at 0 V for various IDT periods. The fitted intrinsic quality factor $Q_{i}$ and linewidth $\gamma_{i}$ are plotted in Fig. \ref{Fundamental}b.
Interestingly, it is found that the fundamental Rayleigh mode has an intrinsic Q that is more than four times higher than that of the periodically poled higher order modes. We attribute the higher $Q_{i}$ to the much less bulk mode coupling as illustrated in the mode profile in Fig. \ref{Fundamental}c. Compared to the Sezawa mode in Fig. \ref{Fig:bulk mode}b, the bulk acoustic radiation is less visible. The better surface mode confinement comes from the lower acoustic velocity of 3600 m/s (at wavelength of 1.8 $\mu$m). Therefore, the low mechanical Q obtained in the main text is not limited by the acoustic loss of the BTO thin film, and a higher Q resonator can be supported by a proper choice of substrate and device design with more complex structure.

\twocolumngrid
\bibliography{BTO_bibliography}

\end{document}